\documentclass[final,5p,times,twocolumn]{elsarticle}

\usepackage[utf8]{inputenc}
\usepackage[T1]{fontenc}
\usepackage{graphicx}
\usepackage{graphbox}
\usepackage{amsmath}
\usepackage{amssymb}
\usepackage{xcolor}
\usepackage{soul}
\usepackage{bm}

\def\bfr{{\bf r}}
\def\bfe{{\bf e}}

\def\bfn{{\bf n}}

\def\bfd{{\bf d}}
\def\bff{{\bf f}}

\def\bfN{{\bf N}}

\def\bfM{{\bf M}}

\journal{Journal}

\begin{document}
\begin{frontmatter}

\title{Mechanics of a Model Frictional Knitted Fabric.}

\author[a]{Laura Michel} 
\author[b]{Antoine Faulconnier}
\author[b]{Audrey Steinberger}
\author[c]{Samuel Poincloux}
\author[a,d]{Jérôme Crassous}

\ead{jerome.crassous@univ-rennes.fr}

\affiliation[a]{organization={PMMH, CNRS, ESPCI Paris, Université PSL, Sorbonne Université, Université de Paris},
            city={Paris},
            postcode={75005}, 
            country={France}}
\affiliation[b]{organization={CNRS, ENS de Lyon, LPENSL, UMR5672},
            city={Lyon cedex 07},
            postcode={69342}, 
            country={France}}
\affiliation[c]{organization={Department of Physical Sciences, Aoyama Gakuin University, 5-10-1 Fuchinobe, Sagamihara, Kanagawa 252-5258},
            country={Japan}}
\affiliation[d]{organization={Université de Rennes},
            city={Rennes},
            postcode={35000}, 
            country={France}}
            
\date{\today}

\begin{abstract}
We present a study of the shapes and mechanical properties of a knitted loop made of elastic and frictional yarns. First, for non-frictional yarns, we investigate how the elastic energy of a loop varies with its shape. This allows us to quantify the energetic contributions of stretching and bending, and to identify energetically favorable loop configurations. We then consider the case of thin frictional yarns. In this case, the stability of a loop can be reduced, even in the presence of friction, to a simple planar \textit{elastica} problem. Friction gives rise to a multiplicity of stable loop shapes in a knitted fabric at rest, \textit{i.e.}, in the absence of externally applied forces. We show that, in the limit of thin yarns, a 2D \textit{elastica} model can predict these rest shapes. Furthermore, the force–deformation curves under uniaxial stretching can be obtained straightforwardly within this framework. Measurements of loop dimensions and forces during uniaxial stretching of nylon knitted fabrics allow us to assess the qualitative and quantitative validity of the planar \textit{elastica} model. Finally, we examine the limitations of this model for dense knitted fabrics.
\end{abstract}

\end{frontmatter}

\section{Introduction}

Knits are yarns entangled into an intricate periodic network of loops, also called stitches. 
Beyond garments we wear everyday, knitted fabrics are heavily used in multiple engineering fields, including composite re-inforcement~\cite{leong2000potential}, biomedical~\cite{zhang2018application}, and emerging applications such as wearable for soft robotics~\cite{sanchez_textile_2021,sanchez20233d}, electricity generation~\cite{chen2020smart}, sensing~\cite{atalay2014weft}, or haptics ~\cite{dupasquier2024haptiknit}. 
While the mechanical aspect of the fabric is fundamental to these applications~\cite{bueno2019structure}, quantitatively understanding how macroscopic responses emerge from mesoscale structural properties and microscale yarn material properties remains an ongoing challenge. 
A quantitative prediction of mechanical properties of knitted fabrics is challenging mainly because of the following two points: (1) the coupling between non-linearities emerging from the looped geometry and those coming from dissipation at the contacts; (2) the constitutive yarn is a slender object, but its relevant lengthscales (yarn diameter, length, radius of curvature) can overlap significantly upon deformation of the fabric. 
Various approaches have been proposed to tackle this challenge, from homogenised formulations~\cite{poincloux.2018b,niu2025geometric} to yarn-based large scale simulations~\cite{kaldor_simulating_2008,liu_computational_2018,ding2024unravelling}. In the following, we consider models that focus on the mechanics of a single unit cell, called a stitch, that assumes bi-periodicity of the knit network. 
This simplified approach places out of the scope changes in stitch topology~\cite{hurd1953fundamental,singal_programming_2024}, finite size rolling effects~\cite{ucar2000prediction,amanatides2022characterizing,tajiri2025curling} or fabric-scale patterning~\cite{yuksel2012stitch,narayanan2018automatic,dupasquier2025multi}.

Early attempts at rationalising knits used geometric descriptions, motivated by the industrial challenge of predicting the area of a fabric given the length of yarn used in its fabrication. 
Peirce introduced a first model to describe the tight limit of Jersey (also called Stockinette) fabrics~\cite{peirce1947geometrical}, which was later refined to avoid torsional discontinuity and include looser fabrics~\cite{leaf1955geometry,munden1959geometry}. 
These models successfully capture that both the row and column densities scale inversely with the length of yarn in a stitch. 
For their simplicity, geometrical models are still routinely used in the textile industry~\cite{lanarolle_geometry_2021}. 
Geometrical models have also been extended to other types of knit patterns~\cite{kurbak2008geometrical,kurbak2009geometrical}, or to elegant bicontinuous surface analogies~\cite{knittel2020modelling}. 
However, geometry alone cannot capture the crucial stress and shape relaxation endured by the fabric after manufacturing~\cite{fletcher1952geometry,knapton1975geometry,allan1983prediction,allan1985prediction}, or the mechanical response to external stresses.

To capture these mechanical effects, models have been developed and usually consist of a 1D elastic rod that reproduces the geometry of a stitch, with boundary conditions imposed by the symmetry of the periodic network. 
Initial approaches focused on the relaxed mechanical state, first relying on ad hoc input for the geometry~\cite{leaf1960models}, but consecutive ones quickly relied on elasticity and constraints alone, either in 2D~\cite{postle1967analysisI} or 3D~\cite{postle1967analysisII,hepworth1976mechanics}. 
These models were then extended to uniaxial~\cite{shanahan1974theoreticalI,shanahan1974theoreticalII,hong2002theoretical} and biaxial~\cite{macrory1975biaxial,hepworth1978biaxial} response to external loads.
The main difficulty and discrepancy among the different models concerns how to handle the effect of yarn-yarn contacts as an external force and torque source. 
In the loose limit where the yarn diameter is much smaller than the length of yarn in the stitch, modelling the contact as an external point force that can travel along the yarn provides quantitative agreement ~\cite{hong2002theoretical,crassous_jersey_2024}.
In the tight limit or under high deformation either in the vertical or lateral directions, most of the yarn is in contact with another one, and the response becomes dominated by yarn-yarn interactions~\cite{gonzalez2025pulling}. 
While most models rely on finding equilibrium shapes by force-torque balance on the elastic rod, others adopt an energetic approach, defining the equilibrium shape as the minimum of elastic energy given external constraints~\cite{dejong1977energyI,dejong1977energyII,choi2003energy,choi2006shape}.
In particular, \cite{choi2003energy} introduces a mapping of the energy landscape in the space drawn by the stitch's lateral and vertical sizes, which we will leverage extensively in this work. 
Finally, these models explore the mechanical response of stitches at low external forces, where bending deformations of the rod dominate the response. 
The other limit is a stitch under strong biaxial tensions, where curvature becomes localized at the contact, and the rest of the yarn remains straight. 
The response then becomes uniquely determined by the geometry of the loop and the stretching response of the yarn~\cite{popper1966theoretical,whitney1966three,karimi2009theoretical,ji2024theoretical}. 
However, apart from early work in a different warp-knit geometry~\cite{kawabata1970method,yanagawa1970experimental,yanagawa1971theoretical}, the transition between the bending and stretching regimes remains poorly understood.
Another overlooked aspect is the role of friction in the mechanical response. 
While in the strong biaxial regime friction simply scales the force response, considering an Euler's capstan contribution at the entanglements~\cite{ji2024theoretical}, in low-stress regimes, friction brings complex hysteretic responses~\cite{dusserre2015modelling} and a degeneracy of the possible rest states~\cite{crassous_jersey_2024}. 
A systematic quantification of the role of friction in the response to external forces remains an open question. 
Beyond the case of single knit stitches, recent advances in numerical, experimental and theoretical modeling of entangled frictional fiber assemblies have unlocked quantitative predictions of their mechanics. First, numerical simulations of rods with frictional contacts has seen important developments~\cite{cirio2015efficient, li2018implicit, crassous_discrete_2023}, now allowing for quantitative predictions \cite{seguin2022twist, crespel2024contact, crassous_jersey_2024, hohnadel2026mixed}. Then, progress in understanding the complex 3D geometry of slender elements in contact~\cite{clauvelin2009matched, grandgeorge2021mechanics, poincloux2021bending, crassous_jersey_2024} has helped reduce complex frictional contact zones to simplified force balances that can be integrated into lower dimensions models. However, quantitative assessments are still lacking to determine when full 3D simulations are necessary and when reduced models are sufficient.

Leveraging both these numerical and theoretical advances, we combine 3D numerical simulations, 2D \textit{elastica} modeling, and experiments to clarify the mechanical behavior of a model knitted stitch. In particular, by quantifying the energy contributions from different deformation modes, we rationalize the transition from bending to stretching regimes and their asymptotic behavior. 
We then characterize how friction affects the set of stable states under constant external forces, and the response of a stitch under uniaxial extensions. We finally detail the relevance and limitations of the 2D model, and how non-local interactions affect the energy landscape as we deviate from the slender limit. 

The paper is organized as follows. After the present introduction, Section 2 details the models we utilize in this study: the 3D numerical model, the 2D analytical model and the experimental one. 
Section 3 is dedicated to the energy landscape configuration in the frictionless and slender limit and discussions on the transition between the bending and stretching-dominated regimes.
The following Section 4 delves into the frictional slender limit.
We first characterize the set of stable stitch configurations at rest or under a constant uniaxial force. 
Then, we discuss uniaxial extensions and, in particular, the role of friction at the contacts. 
Before the conclusion and discussion, the last Section 5 explores the limitations of the 2D model and the consequences of deviating from the slender limit.

\section{Description of the Models.}

\subsection{Hypotheses.}
We describe the elasticity of frictional knitted fabrics using some simplifying hypotheses and restricting to some model geometries. 
We first assume that the fabric is made of a single thread of diameter $d$ that satisfies Kirchhoff rod theory and is extensible. 

The rod is described by its centerline, $\bfr(s):=\big(x(s),y(s),z(s)\big)$, where $s$ denotes the curvilinear abscissa along the rod. The unit tangent to the centerline is defined as $\bfd_3:=d\bfr/ds$. Together with two material vectors $\bfd_1(s)$ and $\bfd_2(s)$, it forms a direct orthonormal basis $(\bfd_1,\bfd_2,\bfd_3)$ attached to the rod. The rotation of this material frame along the rod is described by the twist vector, $\bm{\kappa}=\kappa_1 \bfd_1 + \kappa_2 \bfd_2 + \tau \bfd_3$, such that $d \bfd_i/ds = \bm{\kappa} \times \bfd_i$ for $i=1,2,3$.
The elastic energy of the thread, $\mathcal{E}$, is written as follows:
\begin{equation}
\mathcal{E} =\frac{B}{2}\int \kappa^2 ds  + \frac{C}{2} \int \tau^2 ds + \frac{K}{2} \int \varepsilon^2 ds.
\label{eq:elastic_energy}
\end{equation}
The first term is the bending energy, $\mathcal{E}_{\mathrm{bend}}$, where $B$ is the bending modulus and $\kappa =(\kappa_1^2+ \kappa_2^2)^{1/2}$ is the curvature. 
The second term is the twist energy, $\mathcal{E}_{twist}$, where $C$ is the torsional modulus and $\tau$ is the torsion.  The last term is the stretching energy, $\mathcal{E}_{\mathrm{stretch}}$, where $K$ is the stretching modulus and $\varepsilon=ds/ds_0$ is the relative extension of the centerline. This last term must be included in order to describe stretched configurations of the fabrics (see Section~\ref{sec:stretch}). 
While the  Kirchhoff rod theory corresponds to the limit $K\to\infty$, we consider here an extensible rod made of a perfectly elastic material. 
It follows that $B = EI$, $C = GJ$, and $K = ES$, where $E$ is Young’s modulus and $\nu$ is the Poisson ratio. 
Furthermore, $G=E/2(1+\nu)$, where $I$ denotes the second moment of area, $J$ the torsion constant, and $S$ the cross-sectional area. 
Unless otherwise stated, we take $I = \pi d^4 / 64$, $J = \pi d^4 / 32$ and $S = \pi d^2 / 4$. 
Beyond elasticity, we assume that interactions between threads are frictional and follow Amontons-Coulomb's laws with static and dynamic coefficients both equal to $\mu$.
The geometry investigated in this study is a periodic, quasi-planar knitted fabric submitted to biaxial stress. 
The unit cell of the planar periodic tiling is a rectangular mesh of size $(\ell_x,\ell_y)$. 
In the following, the rest length of the thread per cell is denoted by $\ell$. The mesh structure is quasi-planar in the sense that we constrain the knitted fabric to prevent it from rolling.

\subsection{3D Numerical Model.\label{sec:3D_model}}

\begin{figure}[htbp]
    \centering
    \includegraphics[width=1.\linewidth]{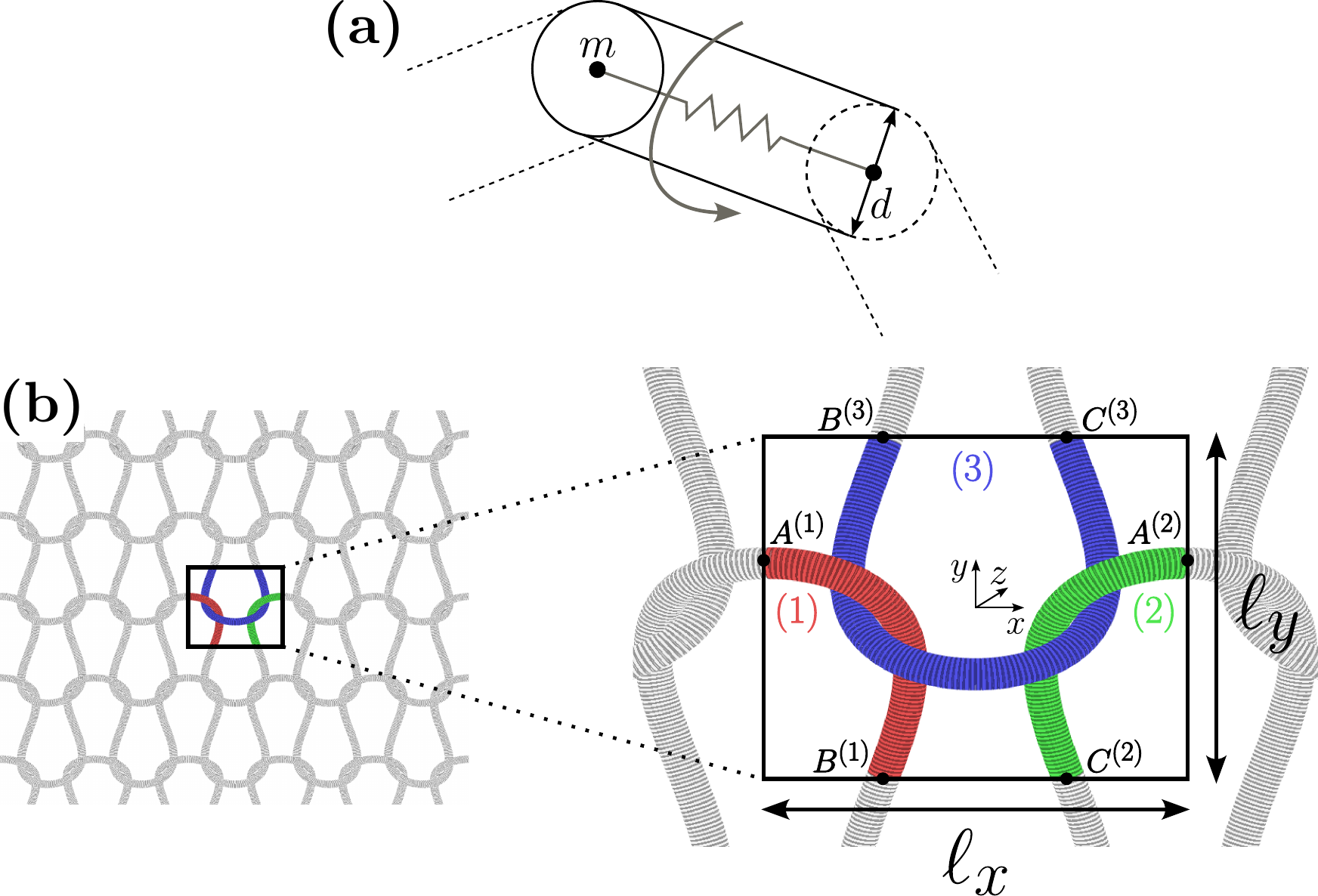}
    \caption{(a) Illustration of a thread segment. (b) Example of a simulated knitted fabric. The fabric is composed of identical rectangular cells of size $\ell_x \times \ell_y$. Each cell comprises three threads (1), (2), and (3). The endpoints of these threads $\{A^{(1)}, B^{(1)}, C^{(2)}, \dots\}$, represented with black dots, are constrained to be on the cell boundaries with periodic boundary conditions.}
    \label{fig:Sim3D}
\end{figure}

In this study, we perform numerical simulations of 3D fabrics that satisfy the hypotheses described in the previous section. 
The fabrics are made of a thread that follows Kirchhoff rod theory and is extensible, such that its elastic energy is described by Eq.~\eqref{eq:elastic_energy}. The simulations are based on a Discrete Elastic Rod (DER) model~\cite{bergou2008discrete}, in which contacts are detected and the contact forces are frictionless, generating only normal components. The approach represents threads as a set of point masses connected by extensional springs (see Fig.~\ref{fig:Sim3D}a). Each spring is associated with a cylinder of diameter $d$, while each mass is located at the center of a sphere of the same diameter, thereby accounting for the thickness of a thread. The framework captures bending, stretching, and twist energies. Indeed, the masses are allowed to translate, thereby accounting for bending and stretching. Furthermore, while the cylinders remain straight, they are allowed to rotate about their axis, enabling the fiber to twist. Contacts between threads are detected using a neighbor-search method. These contacts generate a normal force, modeled as a spring–dashpot system.

In the following, we simulate quasi-planar periodic knitted fabrics in a jersey configuration, similar to that shown in Fig.~\ref{fig:Sim3D}(b). 
A fabric is composed of an arbitrary number of identical rectangular cells of size $\ell_x \times \ell_y$. 
Each cell comprises three threads with total rest length $\ell$, and whose endpoints lie on the cell boundaries. These constraints are enforced through periodic boundary conditions.
In Figure~\ref{fig:Sim3D}(b), point $A^{(2)}$ is a copy of point $A^{(1)}$, translated by $\ell_x \boldsymbol{e_x}$. 
This is ensured by imposing that their positions satisfy $\bfr_{A^{(2)}}-\bfr_{A^{(1)}}=\ell_x \boldsymbol{e_x}$ and that their material frames $(\bfd_1, \bfd_2, \bfd_3)$ are identical.
Similarly, points $B^{(3)}$ and $C^{(3)}$ are the copies of points $B^{(1)}$ and $C^{(2)}$, translated by $\ell_y \boldsymbol{e_y}$. 

Finally, the near-planarity of the fabric is ensured with $\bfr_{B^{(1)}} \cdot \boldsymbol{e_z}=\bfr_{C^{(2)}} \cdot \boldsymbol{e_z}=\bfr_{A^{(1)} \cdot \boldsymbol{e_z}}=0$. Contacts between threads are taken into account within a cell as well as with the eight surrounding cells.
 
\subsection{2D Model of a Thin Jersey.\label{sec:2D_model}}

\begin{figure}[htbp]
   \includegraphics[width=\columnwidth]{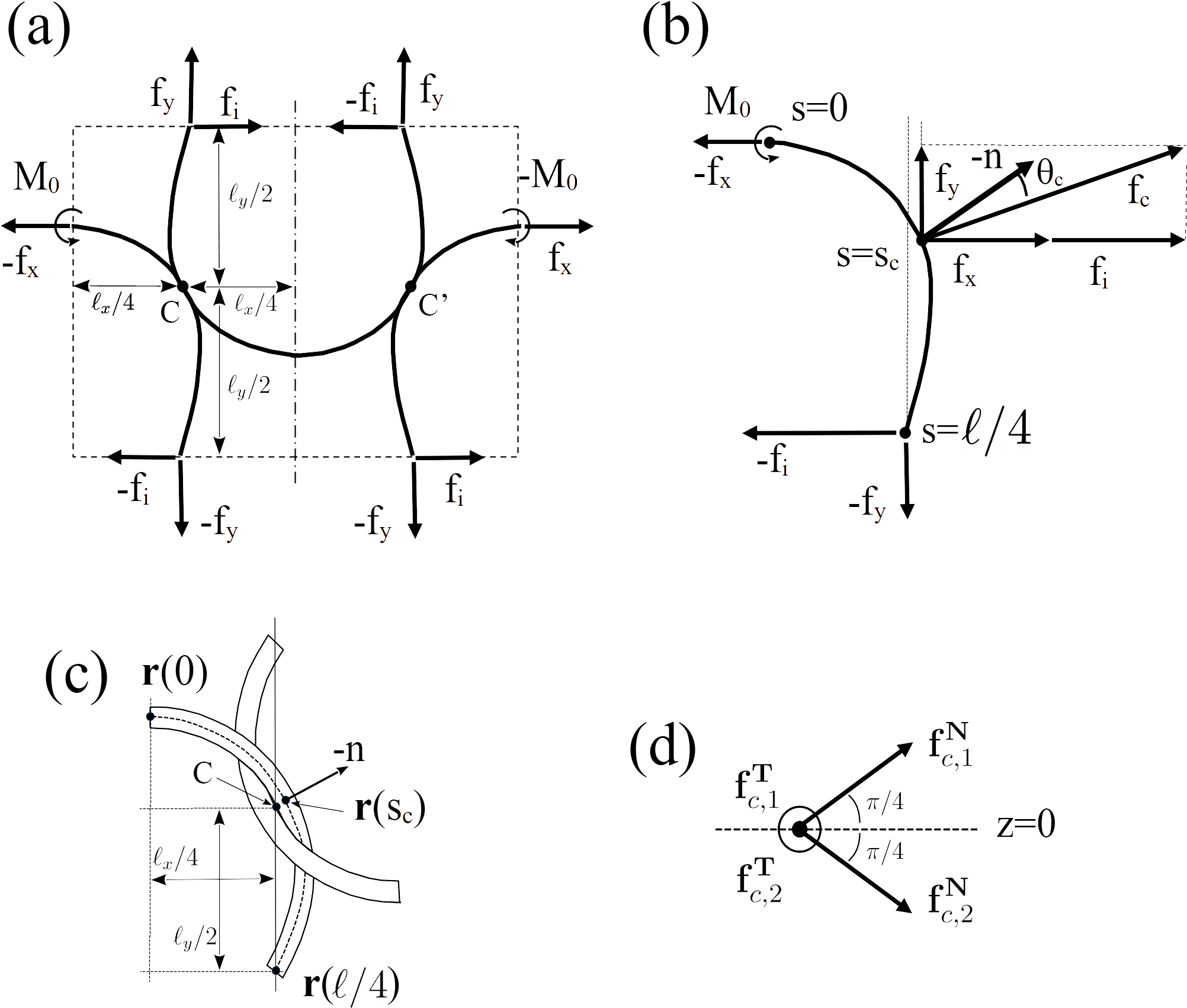}
  \caption{a) Schematic of a mesh made of planar {\it elastica}. $\bff_x$, $\bff_y$, $\bff_i$, and $\bfM_0$ are applied forces and moments. $C$ and $C'$ are the contact points. b) {\it Elastica} forming a quarter of a Jersey mesh of total size $(\ell_x,\ell_y)$. The ends of the {\it elastica} are at $s=0$ and $s=\ell/4$, and the contact force is applied at $s=s_c$. $\bfn$ is the normal vector, and $\theta_c$ ($\theta_c<0$ in the figure) is the angle between $-\bfn$ and the contact force $\bff_c$. c) Position of the point $\bfr(s_c)$ with respect to the $C$ mesh boundaries. d) Orientations of the contact forces $\bff_{c,1}$ and $\bff_{c,2}$ with respect to the $z=0$ plane in the 3D problem. Superscript $N$ (respectively $T$) is for the normal (respectively tangential) component of the contact force.}\label{fig:Elastica2D}
\end{figure}

We also consider a 2D model of knitted fabrics in which threads are treated as 2D \textit{elastica} lying in the plane $z = 0$ and are arranged as shown in Fig.~\ref{fig:Elastica2D}(a). We assume that the mesh has a vertical symmetry plane at $x=\ell_x/2$, which rules out shear deformations of the unit cell. 
The threads touch at points $C$ and $C'$ (see Fig.~\ref{fig:Elastica2D}a), which are axes of rotation of order 2. 
The wallpaper symmetry group of the bidimensionnal stitch is then {\it p2mg}. External forces $\pm \bff_x$, $\pm \bff_y$, $\pm \bff_i$, and external moments $\pm M_0$ are applied to the mesh. 
Since the plane containing the mesh is a plane of symmetry, the forces applied at the ends and at the contacts have no component along $e_z$. 
Similarly, the applied moments are directed along $e_z$. Only one strand is attached to the lateral sides of the mesh, so the external forces have no vertical components, preventing solid rotation of the mesh. In contrast, two strands are attached to the top and bottom sides, allowing horizontal components $\pm \bff_i$ on the strands. These components cancel each other, ensuring no net contribution on the whole side (Fig.~\ref{fig:Elastica2D}a).
Given the {\it p2mg} symmetry, the equilibrium problem of one mesh reduces to that for one quarter of the mesh, as shown in Fig.~\ref{fig:Elastica2D}(b). 
An {\it elastica} of total length $\ell/4$ is fixed at the contact point $C$, located at the curvilinear abscissa $s_c$. 
At the $s=0$ boundary, a moment $\bfM=M_0\bfe_z$ and a force $\bff=-f_x \bfe_x$ are applied. At the other boundary $s=\ell/4$, a force $\bff=-f_i \bfe_x-f_y \bfe_y$ is applied. 

The periodicity of the mesh is obtained by the following conditions on the centerline:
\begin{subequations}
\begin{align}
\frac{dy}{ds}(0)&=0\label{eq::2D::periodicity::x},\\
x(s_c)&=x(\ell/4)-(d/2)\bfn\cdot\bfe_x\label{eq::2D::periodicity::y}.
\end{align}
\label{eq::2D::periodicity}
\end{subequations}
The mesh sizes are related to the positions of the endpoints and the contact point $C$:
\begin{subequations}
\begin{align}
x(s_c)-x(0)&={\ell_x / 4}-(d/2)\bfn\cdot\bfe_x\label{eq::2D::size::x},\\
y(s_c)-y(\ell/4)&={\ell_y / 2}-(d/2)\bfn\cdot\bfe_y.
\label{eq::2D::size::y}
\end{align}
\label{eq::2D::size}
\end{subequations}
At equilibrium, the forces acting on the ends are balanced by the contact force $\bff_c=\bff_x+\bff_y+\bff_i$. 
We define $\theta_c$ as the angle between $\bff_c$ and $-\bfn$, where $\bfn$ is the normal vector to the centerline. An angle $\theta_c\neq0$ is possible only if a tangential force is present at the contact, and hence only in the presence of friction..
For a fixed thread length $\ell$, the different quantities $\ell_x$, $\ell_y$, $f_x$, $f_y$, $f_i$, $\theta_c$, and $s_c$ are not independent. 
They are interrelated by the conditions given by Eqs.~\eqref{eq::2D::periodicity::x}-\eqref{eq::2D::periodicity::y} and Eqs.~\eqref{eq::2D::size::x}-\eqref{eq::2D::size::y}. 
We may consider different problems depending on whether forces $(f_x,f_y)$ or meshes sizes $(\ell_x,\ell_y)$ are given, and whether $s_c$ or $\theta_c$ are given. 
In the following, we consider different kinds of problems depending on which quantities are known or to be determined.
\begin{subequations}
\begin{align}
&\text{Type i)}\qquad & \{\ell_x,\ell_y,s_c\}&\rightarrow \{f_x,f_y,\theta_c\}\qquad\label{problem::type1} \\
&\text{Type ii)}\qquad& \{\ell_x,\ell_y,\theta_c\} &\rightarrow \{f_x,f_y,s_c\}\qquad\label{problem::type2}\\
&\text{Type iii)}\qquad& \{f_x,f_y,s_c\} &\rightarrow  \{\ell_x,\ell_y,\theta_c\}\qquad\label{problem::type3}\\
&\text{Type iv)}\qquad& \{f_x,f_y,\theta_c\} &\rightarrow  \{\ell_x,\ell_y,s_c\}\qquad\label{problem::type4}\\
&\text{Type v)}\qquad& \{f_x,\theta_c,s_c\} &\rightarrow \{\ell_x,\ell_y,f_y\}\qquad\label{problem::type5}
\end{align}
\end{subequations}
For example, for Type i) problems, we fix the abscissa $s_c$ of the contact point, and constrain the endpoint $s=0$ to satisfy Eq.~\eqref{eq::2D::periodicity::x} and Eq.~\eqref{eq::2D::size::x}, and the endpoint $s=\ell/4$ to satisfy Eq.~\eqref{eq::2D::periodicity::y} and Eq.~\eqref{eq::2D::size::y}. 
The {\it elastica} centerline $\bfr(s)$ is then determined, together with the applied forces at the constrained abscissas $s=0$, $s=s_c$, and $s=\ell/4$.
Finally, the quantities $\{f_x,f_y,\theta_c\}$ are obtained. 
For problems of Type iii) or iv), the forces $f_x$ and $f_y$ are applied at the endpoints, while their positions are not constrained by Eq.~\eqref{eq::2D::size::x} and Eq.~\eqref{eq::2D::size::y}. 
These relations are then used to determine the mesh size. 
Problems of Type ii) or iv) are solved for varying $s_c$ until the angle $(-\bfn,\bff_c)$ is equal to the imposed value $\theta_c$.

In the 2D model, the contact between two threads occurs at a single point $C$ (or $C'$). This is a simplified representation of the contact in the 3D problem. 
In a previous study~\cite{crassous_jersey_2024}, we showed that inside the braid, the contact between two threads occurs at two distinct points $C_1$ and $C_2$. 
These two points are separated by a distance $\sim\sqrt{\ell d}$. In the limit of thin threads $d/\ell \ll 1$, this distance satisfies $\sqrt{\ell d}\ll \ell$, and the two contact points may be merged into a single effective contact, with an effective friction coefficient $\mu_{2D} = \sqrt{2}\mu$. Indeed, the two normal contact forces, $\bff^\bfN_{C_1}$ and $\bff^\bfN_{C_2}$, have the same magnitude $f^N$ and form angles of approximately $\pm\pi/4$ with the average plane of the fabric, $z=0$ (see Fig.~\ref{fig:Elastica2D}d). Their projections onto the $z=0$ plane therefore add up to a total normal force of $2 \cos(\pi/4) f^N$.
At the sliding threshold, the two tangential forces have a magnitude $\mu f^N$ and are parallel. The total tangential contact force is then $2 \mu f^N$. 
The ratio between the projections of the total tangential and normal forces onto the $z=0$ plane is then $(2 \mu f^N)/(2 \cos(\pi/4) f^N)=\sqrt{2}\mu$. The resulting effective friction coefficient in 2D is therefore simply given by $\mu_{2D} = \sqrt{2}\mu$.

\subsection{Experimental Model.\label{sec:experimental_model}}
\begin{figure}[htbp]
    \centering
    \includegraphics[width=\columnwidth]{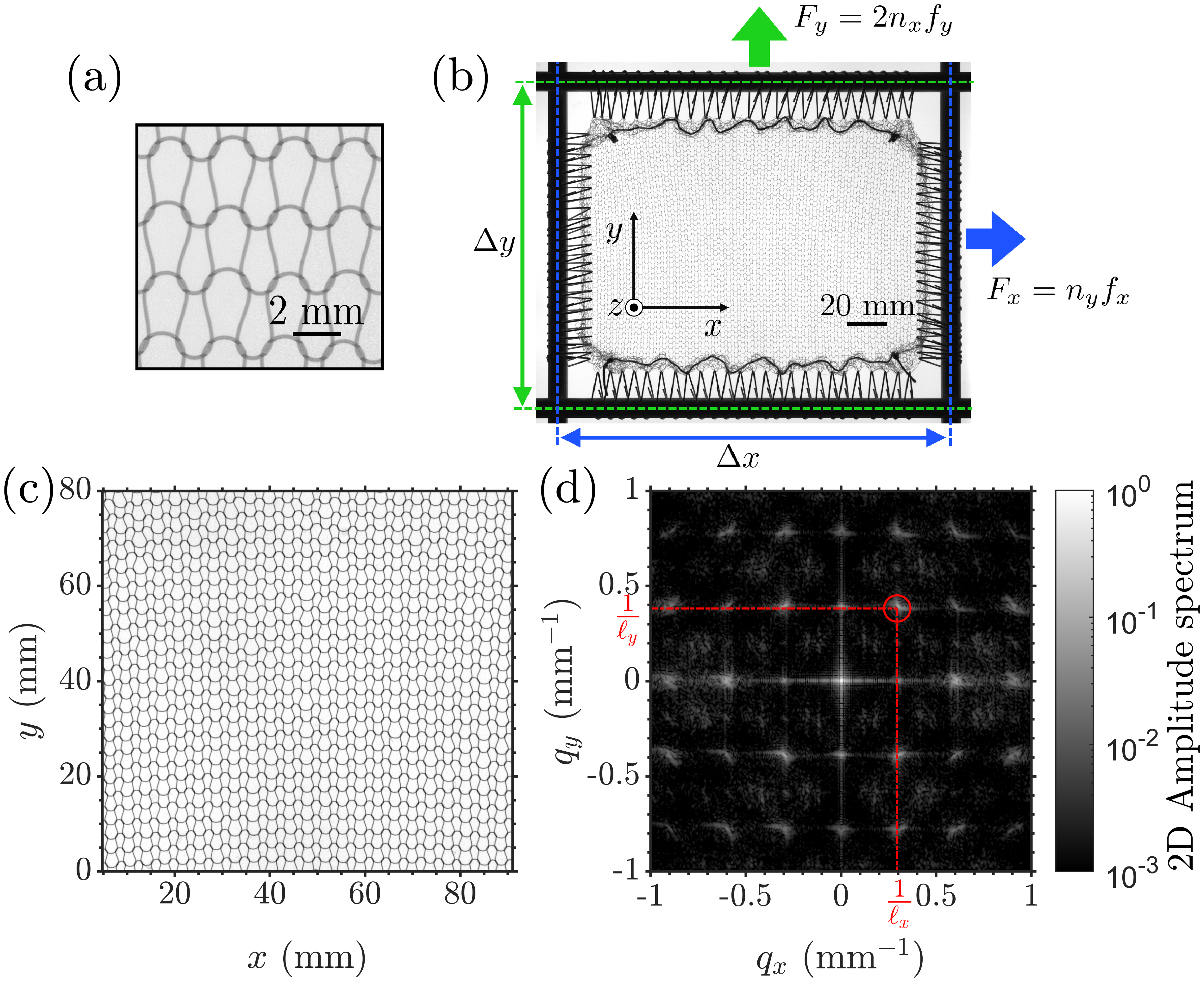}
    \caption{(a) Zoom on the experimental jersey lattice sample. (b) Photograph of the experimental specimen mounted on the frame of the biaxial tensile testing machine. (c) Example of a knit grid used to identify the spatial periodicity of the lattice $(\ell_x, \ell_y)$. (d) Amplitude spectra of the 2D spatial Fourier transform of (c) in the Fourier space, which is used to identify the couple $(\ell_x, \ell_y)$. The position of the fundamental amplitude peak, circled in red, gives the wavenumber associated to the spatial periodicity of the lattice, \textit{i.e.}, at $(q_x, q_y) = (1/\ell_x, 1/\ell_y)$.}
    \label{fig:exp_setup}
\end{figure}

The experimental sample is a model knit made with a cylindrical Nylon filament (Monofil Madeira) as in~\cite{faulconnier2026laddering}, but from a different spool. The thread properties are: diameter $d = 0.140\pm0.002$~mm, linear density $\rho_L = 15.8\pm0.7$ mg m$^{-1}$, Young modulus $E = 1.88\pm0.20$~GPa, bending modulus $B = E\pi d^4/64 =(3.55\pm0.40)\times 10^{-8}$ Nm$^2$, and friction coefficient $\mu = 0.37\pm0.05$. The sample is made of $70 \times 70$ jersey stitches (see Fig.~\ref{fig:exp_setup}(a)), crafted using a knitting machine (Silver Reed SK830).  The length of thread per stitch is $\ell = 9.74\pm0.02$ mm, which corresponds to a loose knit density $d/\ell \approx 1/70$.
The central part of the sample is attached on the frame of a biaxial tensile machine~\cite{crassous_jersey_2024}, as shown in Fig.~\ref{fig:exp_setup}(b), using sliding rounded-shape paper clips. The effective size of the sample is $n_x \times n_y = 50 \times 50$ unit cells, with $n_x$ and $n_y$ designating respectively the number of cells in the $x$- (course) and $y$- (wale) directions. 
The biaxial tensile machine allows us to impose either the spacings $\Delta x$ and $\Delta y$ between the bars, and hence the stretching of the knit sample, or the applied forces $F_x$ and $F_y$ through a feedback loop, in both the $x$- and $y$- directions.
The forces in both directions are measured using strain-gauge sensors, and the positions of the bars are measured using magnetostrictive sensors. In the following, we consider the forces $f_x=F_x/n_y$ and $f_y=F_y/(2n_x)$ acting on the threads at the boundaries of the unit cell.
Moreover, the spatial mesh sizes $\ell_x$ and $\ell_y$ are measured through image processing. For each stretching step of the sample, we record an image of the lattice structure (see Fig.~\ref{fig:exp_setup}(c)) and perform a 2D spatial Fourier transform, as shown in Fig.~\ref{fig:exp_setup}(d). We then identify the coordinates of the fundamental component in Fourier space $(q_x, q_y)$ (see the red circle in Fig.~\ref{fig:exp_setup}d), from which we deduce the spatial periodicities $\ell_x$ and $\ell_y$.

\section{Energy Landscape of a Slender, Frictionless Stitch.}

In this section, we study the elastic energy of a 3D model frictionless fabric ($\mu=0$) made of thin threads, with finite, but small slenderness $d/\ell\ll1$. The simulations, are done with the model presented in Section~\ref{sec:3D_model}, and we assume that the threads are made of a homogeneous and incompressible elastic medium:  $K=E\pi d^2/4$, $B=E\pi d^4/64$, and $C=E \pi d^4/64(1+\nu)$ with $\nu=1/2$.

\subsection{Description of the Energy Landscape.~\label{sec:3D-energy}}

\begin{figure}[htbp]
    \begin{minipage}[c]{1.\linewidth}
        \centering
        \includegraphics[width=1.\linewidth]{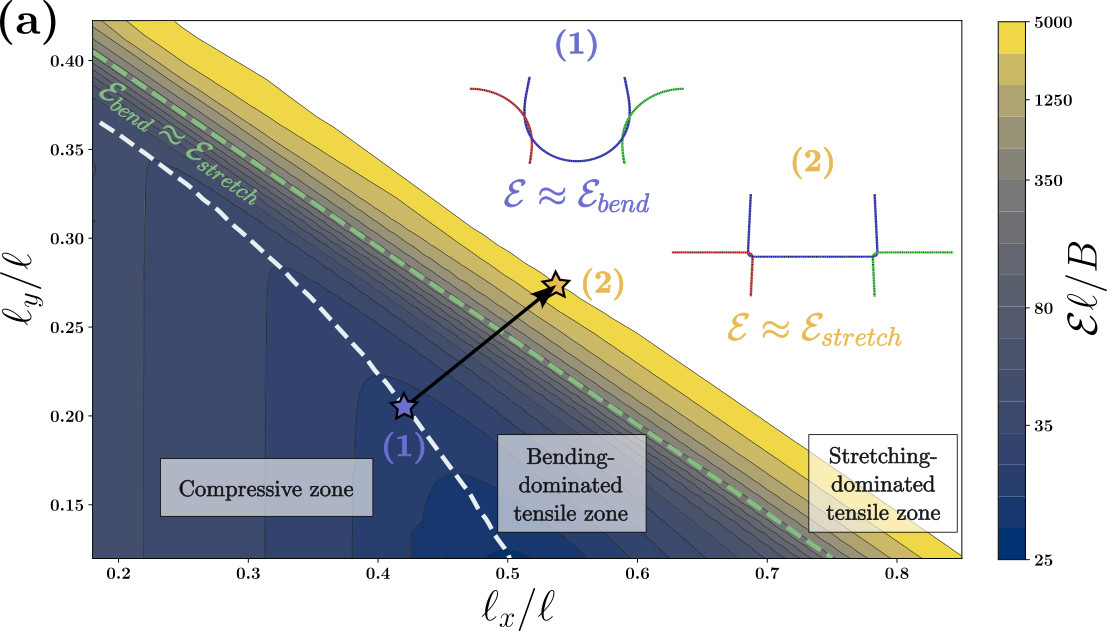}
    \end{minipage}
    \hfill%
    \begin{minipage}[c]{1.\linewidth}
        \centering
        \includegraphics[width=1.\linewidth]{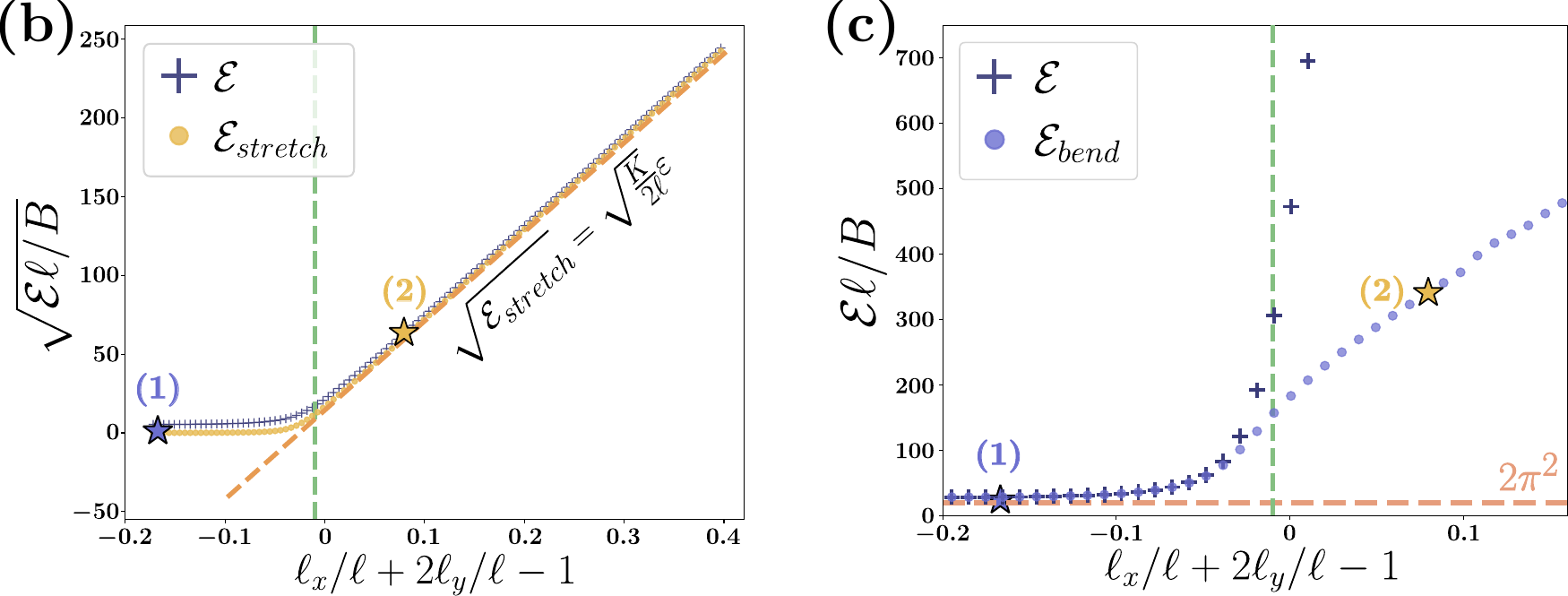}
    \end{minipage}
\caption{(a) Elastic energy $\mathcal{E}$ obtained numerically for a frictionless fabric with $\mu=0$ and $\ell/d = 200$. (White dotted line): Line where $(\partial \mathcal{E}/\partial \ell_x)=0$. (Green dotted line): Line where $\mathcal{E}_{\mathrm{bend}}=\mathcal{E}_{\mathrm{stretch}}$. 
(b) Elastic $\mathcal{E}$ and stretching $\mathcal{E}_{\mathrm{stretch}}$ energies for a fabric initially in configuration (1), progressively stretched beyond configuration (2). 
(c) Elastic $\mathcal{E}$ and bending $\mathcal{E}_{\mathrm{bend}}$ energies for a fabric initially bent below configuration (1), progressively stretched beyond configuration (2).}\label{fig:energy_Landscape}
\end{figure}

We first discuss the variation of the elastic energy with the mesh sizes $(\ell_x, \ell_y)$. 
The mesh sizes are imposed by fixing the endpoints of the three rods to lie on the planes $\{\pm \ell_x/2\}$ and $\{\pm \ell_y/2\}$. 
 
Figure~\ref{fig:energy_Landscape}(a) shows the total elastic energy $\mathcal{E}$ as a function of $(\ell_x, \ell_y)$. 
The energy landscape exhibits a line (shown in white) corresponding to a valley that descends (\textit{i.e.}, the energy decreases) towards large $\ell_x$ and small $\ell_y$. 
It is bordered by regions of increasing energy. 
On the left (respectively, right) of this line, $(\partial \mathcal{E}/\partial \ell_x)_{\ell_y}$ and $(\partial \mathcal{E}/\partial \ell_y)_{\ell_x}$ are negative (respectively, positive): this corresponds to zones where the mesh is compressed (respectively, elongated). The tensile zone is roughly separated into two regions, one dominated by bending energy and the other by stretching energy. Fig.~\ref{fig:energy_Landscape}(a) shows the line where $\mathcal{E}_{\mathrm{bend}}=\mathcal{E}_{\mathrm{stretch}}$ (green dotted line): on the left (respectively, right) of this line, the elastic energy is mainly due to bending (respectively, stretching). 
Note that, for $\mu=0$, there is no tangential component of the contact forces that can exert an axial moment on the thread, and the twist energy is $\mathcal{E}_{\mathrm{twist}}=0$.

\subsection{Stretching-Dominated Configurations.}\label{sec:stretch}

Figure~\ref{fig:energy_Landscape}(b) shows the variations of the elastic energy $\mathcal{E}$ and of the stretching energy $\mathcal{E}_{\mathrm{stretch}}$ along a line joining points labeled (1) and (2) in Fig.~\ref{fig:energy_Landscape}(a). 
The data are plotted as a function of $\ell_x/\ell+2\ell_y/\ell-1$. 
We see that $\mathcal{E} \simeq \mathcal{E}_{\mathrm{stretch}}$ when $\ell_x/\ell+2\ell_y/\ell \gtrsim 1$. For highly stretched threads, the bending energy becomes negligible. As a result, for frictionless threads, the elongation $\varepsilon$ of the thread is constant throughout the thread, \textit{i.e.}, it does not depend on the curvilinear abscissa $s$.
In the high biaxial stretch limit, the stretching energy can thus be written as
\begin{align}
    \mathcal{E}_{\mathrm{stretch}} = \frac{1}{2} K \ell \varepsilon^2,
    \label{eq:label_Estretch}
\end{align}
with the elongation
\begin{align}
    \varepsilon=\frac{\ell_x + 2\ell_y + f(d)}{\ell} -1.
   \label{eq:varepsilon}
\end{align}
In Eq.~\eqref{eq:varepsilon}, $\ell_x + 2\ell_y$ is the length of stretched thread if $d=0$, and $f(d)$ is some excess length of threads due to the geometrical exclusion of thread in the braid zone. It follows that
\begin{align}
    \sqrt{\frac{2 \mathcal{E}_{\mathrm{stretch}}}{K \ell}} = \frac{\ell_x + 2\ell_y + f(d)}{\ell} -1.
    \label{eq:strech_energy}
\end{align}

The function $f(d)$ is determined by the geometry of the threads that curl around each other in the contact zone. If we approximate the threads in these zones as quarter-turns of helices of radius $d/2$ and pitches $2d/\pi$, the excess length is 
\begin{align}
f(d)=2\pi\sqrt{(d/2)^2 + (2d/\pi)^2}\simeq 5d.
\end{align}
The orange dotted line in Fig.~\eqref{fig:energy_Landscape}(b) corresponds to $\bigl(\ell_x + 2\ell_y + f(d)\bigr)/\ell -1$, with $f(d)\simeq 5d$, showing that Eqs.~\eqref{eq:label_Estretch} and~\eqref{eq:varepsilon} (with $f(d)\simeq 5d$) provide a good approximation of the stretching energy.
We checked that Eq.~\eqref{eq:strech_energy}, with $f(d)\simeq 5d$, provides a good approximation for different values of $d$.

\subsection{Bending-Dominated Configurations.}\label{sec:bend}

The bending energy $\mathcal{E}_{\mathrm{bend}}$ for a fabric initially bent below configuration (1), and progressively stretched beyond configuration (2), is shown in Fig.~\ref{fig:energy_Landscape} (c). 

First, we notice that, close to point (1), and more generally in the valley, the values of the dimensionless bending energy are $\mathcal{E}_{\mathrm{bend}}\ell/B\sim 20-30$. 
These values may be understood as follows: near the valley, the threads adopt shapes that are close to circular arcs. 
For a circular thread, the curvature is $\kappa = 2\pi/\ell$, the bending energy is $\mathcal{E}_{\mathrm{bend}}=\frac{1}{2}\int B \kappa^2 ds = 2 \pi^2 B /\ell$, and the dimensionless bending energy is then $\mathcal{E}_{\mathrm{bend}}\ell/B=2\pi^2\approx19.7$. 
This minimum value of the bending energy is close to the values observed in the valley.

Secondly, we notice that, near the valley, the stretching energy is small compared to the bending energy: $\mathcal{E} \simeq \mathcal{E}_{\mathrm{bend}} \gg \mathcal{E}_{\mathrm{stretch}}$. 
Indeed, for a circular loop of radius $\ell/2\pi$, the bending force is $f \sim B (2\pi/\ell)^2$. The stretching energy of a rod of length $\ell$ under a tensile force $f$ is $\mathcal{E}_{\mathrm{stretch}}=f^2 \ell /2 K$. 
For a magnitude of the tensile force which is of the order of the bending force, we obtain $\mathcal{E}_{\mathrm{bend}}/\mathcal{E}_{\mathrm{stretch}} \sim K\ell^2/B\sim (\ell/d)^2\gg 1$ if $\ell/d \gg 1$.

In summary, the elastic energy of a mesh near the valley in the plane of mesh sizes $(\ell_x,\ell_y)$ is close to the bending energy of a circular thread of length $\ell$.

\subsection{Intermediate Configurations.}

\begin{figure}[htbp]
    \centering
    \includegraphics[width=0.9\columnwidth]{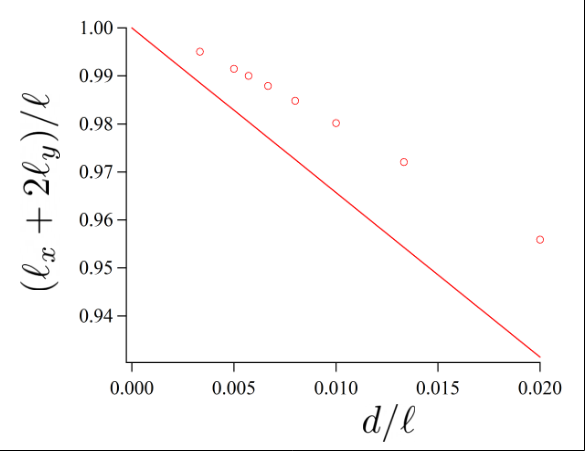}
    \caption{Position of the line $\mathcal{E}_{\mathrm{stretch}}=\mathcal{E}_{\mathrm{bend}}$ as a function of $d$. (Symbols): Values of $\ell_x + 2\ell_y$ for which $\mathcal{E}_{\mathrm{stretch}}=\mathcal{E}_{\mathrm{bend}}$. 3D-Simulations of square meshes $\ell_x=\ell_y$. (Line) Eq.~\eqref{eq:Ebend=Stretch} $\ell_x + 2\ell_y = \ell -3.4 d$.
    } \label{fig:figure_Eb_Es}
\end{figure}

We may divide the mesh-size plane $(\ell_x,\ell_y)$ by a line along which the stretching and bending energies are equal: $\mathcal{E}_{\mathrm{bend}} = \mathcal{E}_{\mathrm{stretch}}$. 
In Fig.~\ref{fig:energy_Landscape}(a), this line is shown as the green dotted line. 
The position of such a line in the plane $(\ell_x,\ell_y)$ can be approximately predicted as follows. 
For highly stretched configurations, we showed in Section \ref{sec:stretch} that $\mathcal{E}\simeq \mathcal{E}_{\mathrm{stretch}}= \frac{K }{2\ell}  \bigl(\ell_x + 2\ell_y + f(d) -\ell\bigr)^2$, with $f(d)\simeq 5d$. 
For configurations with circularly bent threads, $\mathcal{E}\simeq \mathcal{E}_{\mathrm{bend}}\simeq 2 \pi^2 B /\ell$ (see Section~\ref{sec:bend}). 
These two energies are equal when $\bigl(\ell_x + 2\ell_y + f(d)- \ell \bigr)^2 = 4 \pi^2 B /K$, which can be rewritten as
\begin{equation}
    \ell_x + 2\ell_y = \ell - f(d) + \frac {\pi d}{2} \simeq \ell -3.4 d.
    \label{eq:Ebend=Stretch}
\end{equation}
Equation~\eqref{eq:Ebend=Stretch} represents a line in the mesh-size plane $(\ell_x,\ell_y)$. Its proximity to the line along which $\mathcal{E}_{\mathrm{bend}} = \mathcal{E}_{\mathrm{stretch}}$ can be verified as follows.
For square knitted fabrics, $\ell_x=\ell_y$, we measured the values of $\ell_x + 2\ell_y$ for which $\mathcal{E}_{\mathrm{stretch}}=\mathcal{E}_{\mathrm{bend}}$  and compared them to the prediction of Eq.\eqref{eq:Ebend=Stretch}. 
Figure~\ref{fig:figure_Eb_Es} shows the measured values together with the values predicted by Eq.~\eqref{eq:Ebend=Stretch}. One can see that the criterion is reasonably accurate for predicting whether the mesh energy is mainly due to bending or stretching.

\section{Mechanics of a Slender Frictional Stitch.}

As soon as solid friction between threads is taken into account, the description of the mechanical properties in terms of an elastic energy is no longer sufficient.
In particular, the system exhibits metastable states, in the sense that different fabric configurations are possible for a given set of applied external forces. 
In this section, we address the mechanics of slender quasi-planar frictional stitches in the bending-dominated regime, using the two-dimensional model presented in Section~\ref{sec:2D_model}.
The use of this model is justified by the fact that it accurately approximates the elastic energy landscape of the full 3D model in the bending-dominated zone and in the slender limit $d/\ell \ll 1$ (see Section~\ref{sec:comparison_2D_3D}).

\subsection{Lines of Constant Forces.\label{sec:constant_forces}}

In this section, we discuss the following problem: what are all the possible mesh sizes $(\ell_x, \ell_y)$ for given applied forces $(f_x, f_y)$? 
To answer this, we use the 2D \textit{elastica} model presented in Section~\ref{sec:2D_model} and proceeded as follows.
For a given set of applied forces $(f_x, f_y)$, we fixed the abscissa $s_c$ of the contact point and solved the Type iii) problem (Eq.~\eqref{problem::type3}) using $\{f_x, f_y, s_c\}$. 
This gives us the corresponding mesh size $(\ell_x, \ell_y)$ and the contact angle $\theta_c$. 
We repeat this procedure for different values of $s_c$ and finally obtain the entire set of possible mesh sizes as a parametric curve $\bigl(\ell_x(s_c), \ell_y(s_c)\bigr)$ in the configuration space $(\ell_x, \ell_y)$. 
For any set of applied forces $(f_x, f_y)$, there exists a parametric curve $\bigl(\ell_x(s_c), \ell_y(s_c)\bigr)$. 
In this study, we restrict the analysis to two simple cases: the possible mesh sizes for zero applied forces, $(f_x,f_y)=(0,0)$, and for uniaxial applied forces, $f_x=0$ or $f_y=0$.

\subsubsection{Rest-State Line: $(f_x,f_y)=(0,0)$.\label{sec:zeroforce}}

\begin{figure}[htbp]
    \centering
    \includegraphics[width=\columnwidth]{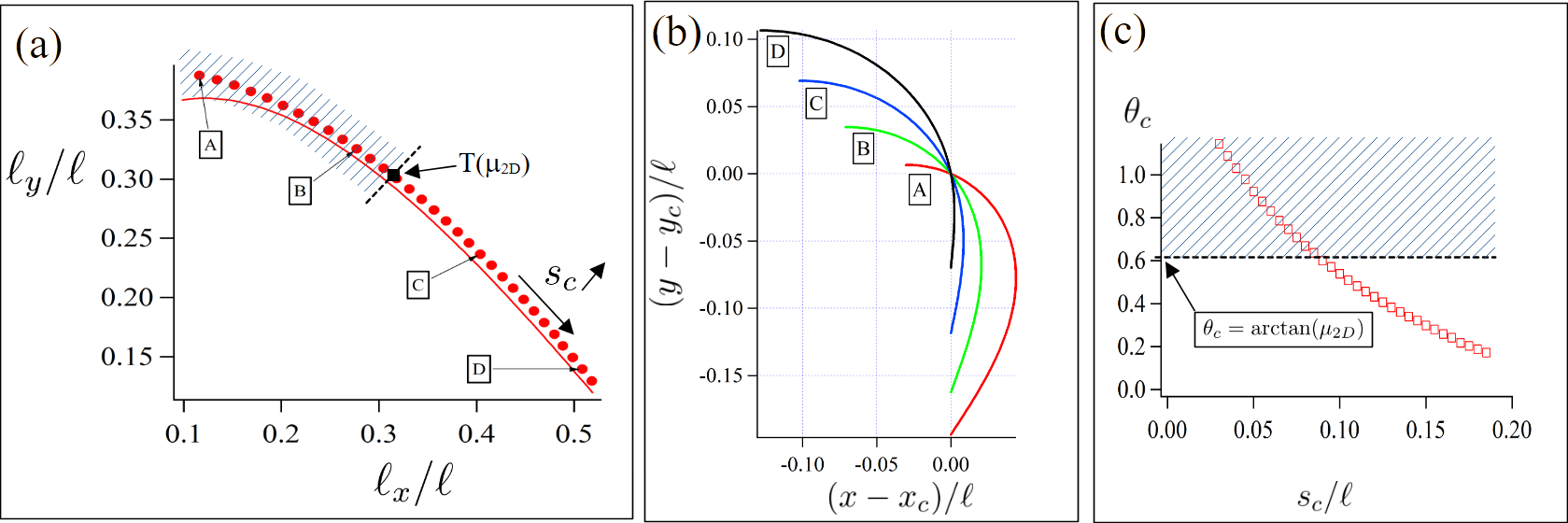}
    \caption{(a) Line of zero applied forces $(f_x,f_y)=(0,0)$. Symbols are results of the 2D model for different $s_c$ from $s_c/\ell=0.03$ (small $\ell_x$, large $\ell_y$) to $s_c/\ell=0.19$ (large $\ell_x$, small $\ell_y$), with steps $\Delta s_c/\ell=0.005$. Letters A to D  identify specific configurations with respective contact arclength $s_c/\ell=0.03$, $0.08$, $0.13$, and $0.18$. The plain line is the analytical prediction of the zero-forces line $(\ell_x^{arc},\ell_y^{arc})$, given by the circular-arc approximation of the {\it elastica}.
    $T(\mu_{2D})$ is the point such that $\tan(\theta_c)=\mu_{2D}$, and the dashed part verifies $\tan(\theta_c) > \mu_{2D}$. 
    (b) {\it Elastica} profiles for the points A, B, C, D of (a).  (c) Variation of $\theta_c$ with $s_c$. 
    The dotted line corresponds to $\theta_c=\arctan(\mu_{2D})$, and the dashed part satisfies $\theta_c > \arctan(\mu_{2D})$.}
    \label{fig:lxly_full}
\end{figure}

Let us consider the case where no external forces are applied at the mesh boundaries. This case has been investigated in a previous study~\cite{crassous_jersey_2024}. We recall here the main results.

Since the forces acting on the threads at the boundaries $x=\pm \ell_x/2$ are zero, mesh equilibrium implies that $f_x = 0$. 
The sum of the forces acting on the threads at the boundaries $y=\pm \ell_y/2$ is also zero. 
Using the symmetry of the mesh with respect to the $y=0$ axis, it follows that $f_y = 0$ (but $f_i \neq 0$). 
This implies that the contact force simplifies to $\bff_c=(f_x+f_i)\bfe_x+f_y\bfe_y=f_i\bfe_x$, with only $f_i$ along the $x$-direction. 

The line of imposed forces $(f_x,f_y)=(0,0)$ is shown in Fig.~\ref{fig:lxly_full}(a). 
Each point on this line corresponds to an {\it elastica} solution for a given value of $s_c$, and all solutions satisfy $\bff_c=f_i\bfe_x$. 
Examples of {\it elastica} profiles are shown in Fig.~\ref{fig:lxly_full}(b) for some mesh sizes $(\ell_x,\ell_y)$ labelled $A-B-C-D$. 
For meshes elongated in the $x$-direction (large $\ell_x$, small $\ell_y$; profile $D$ in Fig.~\ref{fig:lxly_full}b), the contact force, which is oriented along $\bfe_x$, is almost perpendicular to the thread. 
The angle $\theta_c=(\bfn; \bfe_x)$ between the normal vector to the centerline $\bfn$ and the contact force $\bff_c$ is close to zero. 
Such configurations satisfy Coulomb's condition, $\tan(\theta_c) \leq \mu_{2D}$, even for small values of $\mu$; \textit{i.e.}, they are accessible for low friction coefficients. 
By contrast, for meshes elongated in the $y$-direction (small $\ell_x$, large $\ell_y$; profile A in Fig.~\ref{fig:lxly_full}(b)), the angle $\theta_c=(-\bfn; \bfe_x)$ is close to $\pi/2$. 
Such configurations can only be attained for large values of $\mu$.

The evolution of $\theta_c$ with $s_c$ is shown in Fig.~\ref{fig:lxly_full}(c). 
For a given value of $\mu$, only the solutions that satisfy the Coulomb condition $\tan(\theta_c) \leq \mu_{2D}$ are statically admissible. 
Let $T(\mu_{2D})$ be the point on the parametric curve $\bigl(\ell_x(s_c), \ell_y(s_c)\bigr)$ that satisfies $\theta_c(s_c)= \arctan{(\mu_{2D})}$ (see Fig.~\ref{fig:lxly_full}a). 
The part of the curve below this point admits stable solutions without external forces, with friction only balancing the deformed threads. 
In contrast, the part of the curve above $T(\mu_{2D})$ does not satisfy the Coulomb condition and cannot be stabilized by friction alone.
It should be noted that the shape of the parametric curve $\bigl(\ell_x(s_c), \ell_y(s_c)\bigr)$ in Fig.~\ref{fig:lxly_full}(a) does not depend on $\mu_{2D}$; $\mu_{2D}$ only defines the accessible part of this curve.

\subsubsection{Lines of Constant Applied Forces.\label{sec:fixedforces}}

\begin{figure}[htbp]
    \centering
    \includegraphics[width=\columnwidth]{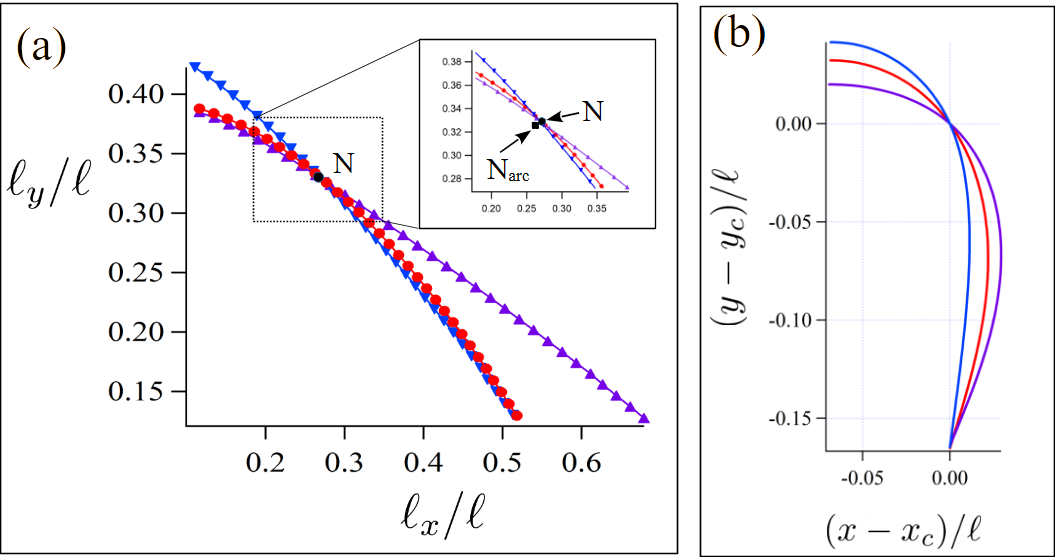}
    \caption{((a) Lines of constant applied forces $(f_x,f_y)$. Symbols correspond to different values of $s_c$ from $s_c/\ell=0.03$ (small $\ell_x$, large $\ell_y$) to $s_c/\ell=0.19$ (large $\ell_x$, small $\ell_y$), with steps $\Delta s_c/\ell=0.005$. Red circles: $(f^*_x,f^*_y)=(0,0)$; purple up triangles: $(f^*_x,f^*_y)=(0,400)$; blue down triangles: $(f^*_x,f^*_y)=(400,0)$. (Black circle): Neutral point $N$. (Black square): Neutral point $N_{arc}$ calculated with the circular-arc approximation. (b) Profiles of the {\it elastica} at the neutral point $N$. Blue line: $(f^*_x,f^*_y)=(0,400)$; red line $(f^*_x,f^*_y)=(0,0)$; purple line $(f^*_x,f^*_y)=(400,0)$.}
    \label{fig:lxly_fixed_forces}
\end{figure}

The results obtained for the zero-force case, $(f_x,f_y)=(0,0)$, may be extended to arbitrary constant forces $(f_x,f_y)$. 
In this study, we focus on the simple cases of uniaxial forces.
Figure~\ref{fig:lxly_fixed_forces}(a) shows the lines obtained for zero forces $(f_x,f_y)=(0,0)$, for uniaxial forces in the $x$-direction $(f^*_x,f^*_y)=(400,0)$, and for uniaxial forces in the $y$-direction $(f^*_x,f^*_y)=(0,400)$. 
In each case, the possible mesh sizes form parametric curves in the configuration space. 
Remarkably, these three lines (and more generally any curve such that $(f_x,f_y)=(0,f_y)$ or $(f_x,0)$) approximately intersect at a single point, which we name the neutral point $N$. 
The corresponding mesh size is then an equilibrium shape for all jersey meshes subjected to forces applied only along $x$ (\textit{i.e.}, $f_y=0$), or only along $y$ (\textit{i.e.}, $f_x=0$), provided that friction is high enough for $N$ to be below $T(\mu_{2D})$.
Figure~\ref{fig:lxly_fixed_forces}(b) shows three profiles obtained at point $N$ for $(f^*_x,f^*_y)=(0,0)$, $(400,0)$, and $(0,400)$. 
We see that the positions of the contact point, $s_c$, differ among the three profiles. 
The computed bending energies are also different, the lowest one being that associated with $(f_x,f_y)=(0,0)$. 
This point has the remarkable property that a mesh of the corresponding size is {\it indifferent} to the application of uniaxial forces, in the sense that uniaxial forces can be applied without any change in mesh size. In the next section, we further investigate the interpretation of the neutral point $N$ by introducing an approximation of the {\it elastica} model that allows us to capture its properties analytically.

\subsection{Circular-Arc Model for the Rest-State Curve $(f_x,f_y)=(0,0)$ and the Neutral Point $N$.\label{sec:arc:model}}

\begin{figure}[htbp]
    \centering
    \includegraphics[width=0.8\linewidth]{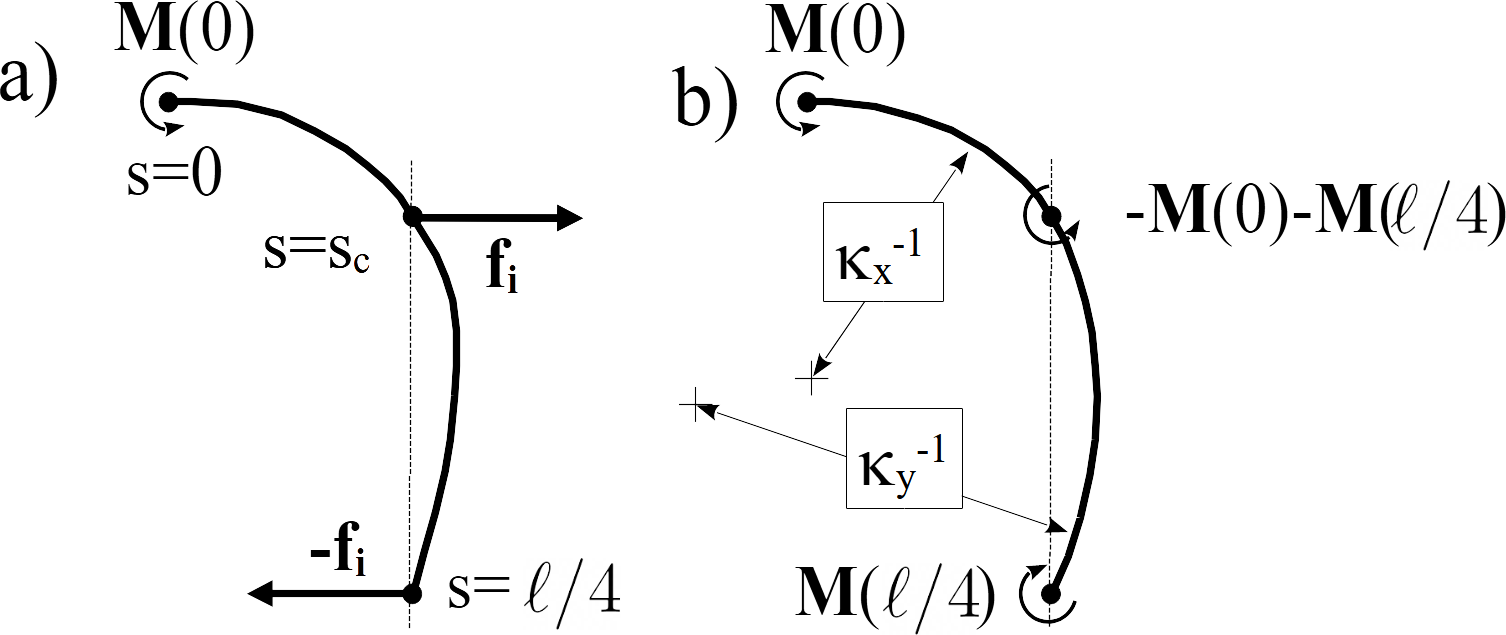}
    \caption{(a) System of forces and moments acting on the {\it elastica} for a mesh at rest. (b) Approximate system of moments. The {\it elastica} is composed of two circular arcs with radii $\kappa_x^{-1}$ and $\kappa_y^{-1}$.}
    \label{fig:forces_vs_moments}
\end{figure}

In the preceding sections, we demonstrated {\it i}) the existence of a family of unit cells satisfying $(f_x,f_y)=(0,0)$, which can be represented as a curve in the mesh-size plane $(\ell_x, \ell_y)$, and {\it ii}) that all the curves obtained for $(f_x,f_y)=(f_x,0)$ and $(f_x,f_y)=(0,f_y)$ intersect at a particular point, which we call the neutral point $N$. 
In this section, we propose a description of the rest-state curve and an interpretation of the neutral point by introducing a simplified model of the \textit{elastica} composing the mesh. 
The simplification is based on the similarity between the \textit{elastica} shape obtained by applying a moment $\bfM(0)$ and a couple of forces $\pm\bff_i$ (as represented in Fig.~\ref{fig:forces_vs_moments}(a)), and that obtained by applying moments $\bfM(0)$ and $\bfM(\ell/4)$ at the extremities, and moment $\bfM(s_c)=-\bfM(0)-\bfM(\ell/4)$ at $s=s_c$  (see Fig.~\ref{fig:forces_vs_moments}(b)).
Within this approximation, the \textit{elastica} is composed of circular arcs of constant curvatures:
\begin{subequations}
\begin{align}
\kappa^{arc}(s) &=\kappa_x,&\qquad &0 \le s < s_c\label{eq:curvature:kappax}\\
\kappa^{arc}(s) &=\kappa_y,&\qquad &s_c < s \le \ell/4\label{eq:curvature:kappay}
\end{align}
\end{subequations}
where $\kappa_x$ and $\kappa_y$ are two constants. Eq.~\eqref{eq:curvature:kappay} may be understood as follows: although the curvature of the \textit{elastica} in Fig.~\ref{fig:forces_vs_moments}(a) decreases between $s_c$ and $\ell/4$, from $\kappa(s_c)=M(0)/B$ to $\kappa(\ell/4)=0$, we ignore these variations in this model and treat $\kappa$ as constant.
With this approximation, the bending energy becomes a function of the parameters $\kappa_x$, $\kappa_y$, and $s_c$:
\begin{equation}
\mathcal{E}^{arc}_{bend}(\kappa_x,\kappa_y,s_c),
\end{equation}
instead of a functional $\mathcal{E}_{\mathrm{bend}}(\kappa(s),s_c)$ of the curvature field $\kappa(s)$, and $s_c$. 

Let us first propose a prediction for the rest-state line using our simplified circular-arc model. For a given value of $s_c$, we seek the values of $\kappa_x$ and $\kappa_y$ that minimize the bending energy, subject to the constraints imposed by {\it i})  the periodicity of the knitted structure, {\it ii}) the continuity of $d\mathbf{r}/ds$ at $s=s_c$, and {\it iii}) the external moments applied at $s=0$, $s=s_c$, and $s=\ell/4$. 
The constrained minimization problem is described in the Appendix.
When the forces are zero, $(f_x,f_y)=(0,0)$, the constant curvatures in Eqs.~\eqref{eq:curvature:kappax} and~\eqref{eq:curvature:kappay} satisfy $\kappa_x=2\kappa_y$. 
Computing the associated mesh sizes $\ell_x$ and $\ell_y$ (see Appendix) gives
\begin{subequations}
\begin{align}
    \ell_x^{arc}(s_c)&=\frac{(6 s_c +\ell/2)}{\pi}\sin\Bigl(\frac{\pi s_c}{3s_c/2+\ell/8}\Bigr), \label{eq:ellx0} \\
    \ell_y^{arc}(s_c)&=\frac{2(6 s_c +\ell/2)}{\pi}\cos\Bigl(\frac{\pi s_c}{3s_c/2+\ell/8}\Bigr). \label{eq:elly0} 
\end{align}
\end{subequations}
These predictions are shown as a line in Fig.~\ref{fig:lxly_full}(a). The predicted rest-state line closely follows the line computed numerically from the full \textit{elastica} equations in the absence of external forces, without any adjustable parameter. 

Let us now use the simplified circular-arc model to interpret and approximately predict the position of the neutral point $N$.
The previous result relied on the assumption that the forces were null, $(f_x,f_y)=(0,0)$. If these forces are non-zero, the relation $\kappa_x=2\kappa_y$ is no longer valid. 
However, the geometric relations linking the mesh sizes $\ell_x$ and $\ell_y$ to the values of $\kappa_x$, $\kappa_y$, and $s_c$ remain valid. 
By examining the variations of the mesh sizes $\Delta \ell_x$ and $\Delta \ell_y$ with respect to variations in $\Delta \kappa_x$, $\Delta \kappa_y$, and $\Delta s_c$, it can be shown that the mesh sizes can remain unchanged ($\Delta \ell_x = \Delta \ell_y =0$) for finite variations of the curvatures around a particular value of $s_c$ on the curve defined by Eqs.~\eqref{eq:ellx0}-\eqref{eq:elly0}. 
The computation is presented in the Appendix, and the solution is:
\begin{subequations}
\begin{align}
s_c^N/\ell &= 0.0740..\\
\ell_x^N/\ell &= 0.2504..\\
\ell_y^N/\ell &= 0.3322..
\end{align}
\end{subequations}
Without any adjustable parameter, the neutral point $N^{\mathrm{arc}}$ calculated using the circular-arc model is shown in the inset of Fig.~\ref{fig:lxly_fixed_forces}(a). It lies very close to the numerical estimate. 
We interpret the neutral point $N$ as follows. 
It corresponds to the mesh size for which, while preserving the constraints imposed by the periodicity of the knit structure, a displacement of the contact point can accommodate a variation in curvature of the {\it elastica}, while maintaining a constant mesh size.

\subsection{Uniaxial Extensions.\label{sec:constant_mu}}

We now address the problem of describing the force and deformation of a fabric during a uniaxial extension in either the $x$- or $y$-direction.
In the presence of friction, the response of a stitch depends on its loading history, and in particular on the sign of the contact angle $\theta_c$. 
In the following, we assume that the initial configurations have $\theta_c>0$.

\begin{figure}[htbp]
    \centering
    \includegraphics[width=1.0\linewidth]{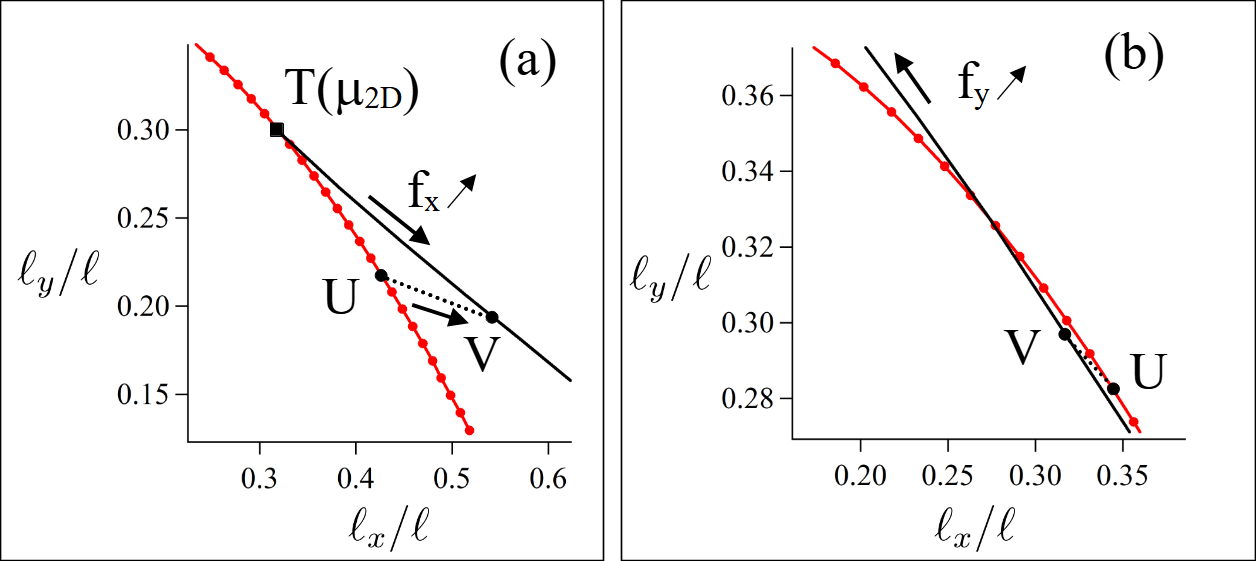}
    \caption{(a) Load trajectory for a $x$-uniaxial force. 
    (Plain black line): line of meshes satisfying $(f_y,\theta_c)=(0,\arctan{(\mu_{2D})})$. 
    (Red circles): line of meshes with zero forces $(f_x, f_y)=(0,0)$. 
    (Dotted black line): a no-sliding trajectory (constant $s_c$) joining the line $(f_x, f_y)=(0,0)$ to the line $(f_y,\theta_c)=(0,\arctan{(\mu_{2D})})$. 
    At point $T(\mu_{2D})$, $f_x=0$ increases as $\ell_x$ increases. $\theta_c$ increases from $\theta_c < \arctan{(\mu_{2D})}$ at point $U$ to $\theta_c = \arctan{(\mu_{2D})}$ at point $V$. 
    (b) Load trajectory for a $y$-uniaxial force. 
    (Plain black line): line of meshes satisfying $(f_y,\theta_c)=(0,-\arctan{(\mu_{2D})})$. 
    (Red circles): line of meshes with zero forces $(f_x, f_y)=(0,0)$. 
    (Dotted black line): a no-sliding trajectory (constant $s_c$) joining the line $(f_x, f_y)=(0,0)$ to the $(f_y=0, \tan{\theta_c} = -\mu_{2D})$ line. 
    $\theta_c$ decreases and changes sign from $0<\theta_c < \arctan{(\mu_{2D})}$ at point $U$, to $\theta_c = -\arctan{(\mu_{2D})}$ at point $V$.
    }\label{fig:fixed_theta}
\end{figure}

We first consider the case of a uniaxial extension in the $x$-direction, where the initial state is a fabric at rest (\textit{i.e.}, no applied stress, $(f_x, f_y)=(0,0)$). 
The initial mesh size $(\ell_x,\ell_y)$ thus lies on the parametric curve $\bigl(\ell_x(s_c), \ell_y(s_c)\bigr)$ discussed in Section~\ref{sec:zeroforce}, with  $\theta_c \le \arctan{(\mu_{2D})}$. 
Let us first consider the case where the initial state is at point $T(\mu_{2D})$. 
When $f_x$ increases, the mesh size elongates in the $x$-direction, and $s_c$ must increase. 
This displacement of the contact point implies that the contact force lies on the Coulomb cone: $\vert \tan{\theta_c} \vert = \mu_{2D}$. 
The trajectory must then satisfy $f_y=0$ and $\theta_c=\arctan{(\mu_{2D})}$. 
This trajectory may be calculated as a solution to the 2D~{\it elastica} problem of Type iv) (Eq.~\eqref{problem::type4}). For a given value of $f_x$, we compute the solution of Eq.~\eqref{problem::type4}, and obtain a mesh size $(\ell_x, \ell_y)$. 
We repeat this procedure for different values of $f_x$, and obtain the meshes verifying $\{f_x\ge 0, f_y=0, \theta_c=\arctan{(\mu_{2D})}\}$ as a parametric curve in the configuration space (Fig.~\ref{fig:fixed_theta}(a), black line).
When $f_x$ increases, the mesh runs down this curve, which always satisfies $\theta_c=\arctan{(\mu_{2D})}$. 
This curve is the trajectory followed by a knitted fabric under a uniaxial extension in the $x$-direction. 

Let us now consider the case of an initial state at rest which is below point $T(\mu_{2D})$ (for instance point $U$ in Fig.~\ref{fig:fixed_theta}(a)). 
In this case, the initial contact angle $\theta_c$ is smaller than $\arctan{(\mu_{2D})}$, and the friction force is inside the Coulomb cone.
During the initial stage of extension, the friction force remains inside the Coulomb cone, and the contact does not slide. 
The abscissa $s_c$ of the contact is constant, and the angle $\theta_c$ increases (dotted line in Fig.~\ref{fig:fixed_theta}(a), given by the solutions of the Type iii) problem (Eq.~\eqref{problem::type3}).
When the line $\theta_c = \arctan(\mu_{2D})$ is reached (point V), the friction force is on the Coulomb cone, and stitches deform at fixed $\theta_c$. 

The discussion for the case of a uniaxial extension in the $y$-direction is quite similar. We can define a parametric curve for meshes satisfying $\{f_x=0,f_y > 0,~\theta_c=-\arctan{(\mu_{2D})}\}$, where $f_y$ is the varying parameter. 
The sliding direction of $s_c$ being opposite to the previous case, the frictional force must also inverse its direction, hence $\theta_c=-\arctan{(\mu_{2D})}$. 
This curve is shown in Fig.~\ref{fig:fixed_theta}(b). 
Starting from a fabric at rest at point $U$, the fabric first deforms with fixed $s_c$ until the contact force reaches the Coulomb condition $\tan{\theta_c} = -\mu_{2D}$ at point $V$. 
During this phase, $\theta_c$, which is initially positive, changes sign and becomes negative. 
When point $V$ is reached, the mesh follows the curve defined by $\{f_x=0,f_y > 0,~ \theta_c=-\arctan{(\mu_{2D})}\}$.

\subsection{Comparison of the 2D Model with Experiments.\label{sec:compa_experience}}

\begin{figure}[h]
    \centering
    \includegraphics[width=1.0\linewidth]{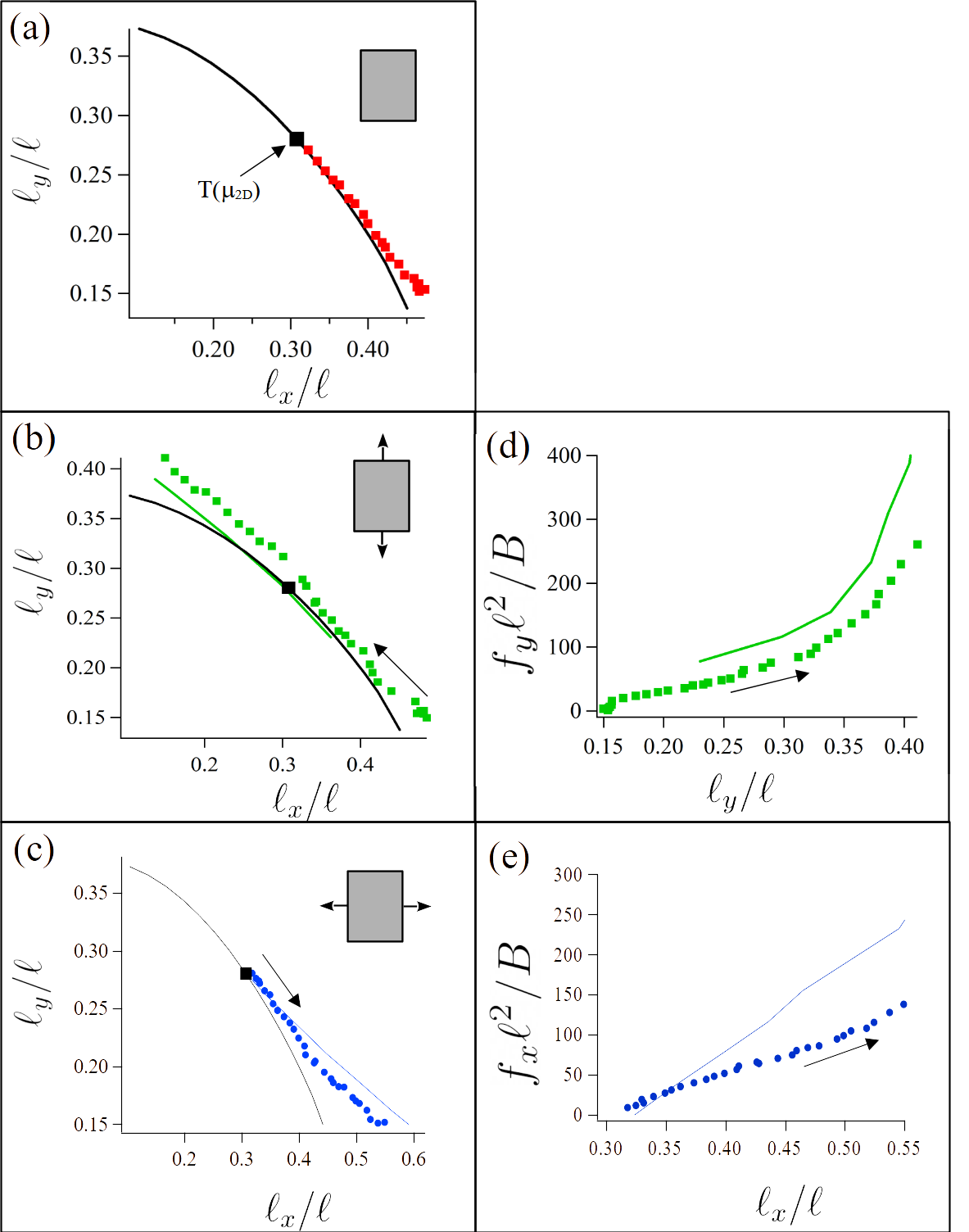}
   \caption{Comparison between the 2D model and experiments. 
   (a-b-c): (Colored symbols): experimental meshes size for (a) knitted fabrics at rest, (b) $y$-uniaxial extension, and (c) $x$-uniaxial extension.
   (Colored lines): trajectories in the $(\ell_x,\ell_y)$ plane for the 2D model fabrics. 
   (Black lines): line of rest states $(f_x,f_y)=(0,0)$ for the 2D model fabrics. 
   (Black squares): terminal point $T(\mu_{2D})$. 
   (Grayed inset): schematic of the type of experiment. (d-e): (Colored symbols): forces in the $y$- (d) and $x$- (e) directions during the uniaxial extensions. 
   (Colored lines): forces for the 2D model fabrics. 
   The parameters used for the 2D model are $\mu_{2D}=\sqrt{2}\mu$, with $\mu=0.35$, and $d/\ell=1/70$.} 
\label{fig:compa_experience}
\end{figure}

We now compare the results of the 2D model of a single thread with one contact point with laboratory experiments.
For this purpose, we used the experimental model detailed in~\ref{sec:experimental_model}, and we performed different experiments: we measured the mesh sizes of fabrics at rest (zero force) and performed uniaxial extensions in the $x$- and $y$-directions.

\subsubsection{Experimental Rest States: $(f_x,f_y)=(0,0)$.\label{subsec:exp_rest_state}}
 
The 2D model predicts that, for given applied external forces $(f_x, f_y)$, multiple mesh sizes are possible, which lie on a parametric curve $\bigl(\ell_x(s_c),\ell_y(s_c)\bigr)$.
We discuss here the particular case of rest states corresponding to zero applied forces $(f_x,f_y)=(0,0)$.
The experiments are performed as follows.

We start around the terminal point, \textit{i.e.}, the largest value of $\ell_y/\ell$ such that $(f_x,f_y)=(0,0)$ (we should note that, in practice, the zero of force is slightly shifted experimentally, because a minimum tension $f_\mathrm{min}^* \approx 24$ is required to maintain the sliders under tension, compensating the effect of gravity). 
From this point, {\it i}) the fabric is stretched in the $x$-direction until reaching an arbitrary non-negligible tension $f_x^* \approx 130$, at which {\it ii}) the fabric is contracted in the $y$-direction by a step $\delta y$. 
Next, {\it iii}) the tension in the $x$-direction is decreased, using a feedback loop, until returning to $f_x^* \approx f_\mathrm{min}^*$, reaching a new point on the line of rest states. At this new position, an image of the fabric is recorded to measure $\ell_x$ and $\ell_y$.

The repetition of steps {\it i}) to {\it iii}) allows us to scan the line of rest states from top (high $\ell_y$) to bottom (low $\ell_y$), until reaching the displacement range limitation ($\ell_y/\ell \approx 0.15$) of the biaxial device. 
The different rest states obtained through this protocol are shown as symbols in Fig.~\ref{fig:compa_experience}(a). 
We see that there is a multiplicity of rest states, and that the possible mesh sizes are located on a line in the $\bigl(\ell_x,\ell_y\bigr)$ plane. 
The location of this line is quite close to the line of rest states obtained from the 2D model (black line in Fig.~\ref{fig:compa_experience}(a)). 
We saw in Section~\ref{sec:zeroforce} and Fig.~\ref{fig:lxly_fixed_forces} that the mesh sizes on the upper part of this curve do not satisfy the Coulomb condition $\tan{(\theta_c)}\le\mu_{2D}$, such that all points above $T(\mu_{2D})$ are forbidden. 
This is in agreement with what is observed experimentally: all possible experimental configurations are located below point $T(\mu_{2D})$, which lies very close to the mesh size with the smallest $\ell_x$ -- largest $\ell_y$ that can be attained experimentally at rest.

\subsubsection{Experimental Uniaxial Extensions. \label{subsec:exp_extension}}

We now turn to the experimental uniaxial extensions of the fabric. 
The $y$-uniaxial extension experiment is performed as follows. 
We first elongate the fabric at a maximal value $\ell_x/\ell\simeq0.48$ in the $x$-direction and adjust $\ell_y$ to bring forces $f_x$ and $f_y$ to $0$. 
Then the fabric is elongated in the $y$-direction, and $\ell_x$ is adjusted
to keep $f_x=0$ (in practice, $f_x = f_\mathrm{min}$). The obtained trajectory is represented by symbols in Fig.~\ref{fig:compa_experience}(b). 
All the meshes are located on a line. 
Figure~\ref{fig:compa_experience}(b) also shows the trajectory predicted by the 2D model (green line), which remains close to and parallel to the experimental data.
The force $f_y$ during the extension (recalling that $f_x=0$) is shown in Fig.~\ref{fig:compa_experience}(d) (green symbols), together with the prediction of the 2D model (green line). 

The experiment for the $x$-uniaxial stretch is performed similarly. 
We elongate the fabric at a maximal value $\ell_y/\ell\simeq 0.30$ in the $y$-direction and adjust $\ell_x$ to bring forces $f_x$ and $f_y$ to $0$. 
Then, the fabric is elongated in the $x$-direction and $\ell_y$ is adjusted to keep $f_y=0$. The resulting trajectory is shown as symbols in Fig.~\ref{fig:compa_experience}(c). All the meshes form a line that closely follows the line predicted by the 2D model (blue line in Fig.~\ref{fig:compa_experience}(c)). 
The corresponding force $f_x$ during the extension (recalling that $f_y=0$) is shown in Fig.~\ref{fig:compa_experience}(e) (blue symbols), together with the prediction of the 2D model (blue line).

\subsubsection{Discussion of the Accuracy of the 2D Model Predictions.\label{subsec:exp_discussion}}

For knitted fabrics made with thin yarns, the equilibrium solutions for given values of the forces and contact angle $\{f_x, f_y, \theta_c\}$ can be obtained as solutions to 2D {\it elastica} problems with appropriate boundary conditions. 
It follows that the set of solutions for given forces $(f_x, f_y)$, or for a given force and angle, for example $\bigl(f_x=0, \theta_c=\arctan{(\mu_{2D})}\bigr)$, can be obtained as parametric curves in the mesh-size plane $(\ell_x,\ell_y)$. 
These predicted curves are clearly observed experimentally. Furthermore, the predictions of the 2D model are also quantitatively accurate.
It should be noted that the set of curves $(\ell_x,\ell_y)$ predicted by the 2D model for the different cases (zero forces, elongation in the $y$-direction, elongation in the $x$-direction), as well as the force values $f \ell^2/B$ during the elongations, are obtained without any adjustable parameters: the ratio $\ell /d$, the friction coefficient $\mu_{2D}=\sqrt{2} \mu$, and the bending modulus $B$ are all estimated from independent measurements.
The 2D model not only provides an interpretation of uniaxial tensile experiments, but also provides a quantitatively accurate approximation of the mechanical behavior of knitted fabrics under tensile stress.
The main discrepancies are observed in the forces, in particular for uniaxial elongation experiments along the $x$-direction (Fig.~\ref{fig:compa_experience}(e)).
We believe that such differences stem from plastic deformations undergone by the nylon filament during knitting.
While small, these deformations imprint natural curvatures on the filament, resulting in a non-straight minimal-energy configuration.
Natural curvatures may slightly alter the energy landscape, and in particular the steepness of its slopes.


\section{Limitations of the 2D Model.}

\subsection{Differences and Similarities Between the 2D and 3D Models.}

We have developed two types of models with different levels of description. In this section, we first recall the main differences and similarities between these models.

In both models, we consider only the behavior of a single stitch unit cell and impose the periodicity of the knitted structure. Planarity is exact in the two-dimensional model, whereas it is imposed only at the boundaries of the mesh in the 3D model.
Furthermore, the 2D model assumes a certain symmetry of the mesh (wallpaper group {\it p2mg}), which allows the mechanical behavior of the mesh to be deduced from one quarter of the unit cell. 
Steric exclusion between yarns is rigorously taken into account in the 3D model, whereas it is only approximated in the 2D model. 
The terms $(d/2)\mathbf{n}\cdot\mathbf{e}_x$ and $(d/2) \mathbf{n}\cdot\mathbf{e}_y$ in Eqs.~\eqref{eq::2D::periodicity::y}, \eqref{eq::2D::size::x}, and \eqref{eq::2D::size::y} represent in-plane shifts but do not account for the fact that a certain yarn length is required to leave the plane at the crossing points between yarns within a stitch.

The description of contact between yarns also differs.
In the 3D model, contact between two yarns consists of two contact points involving only normal forces.
In the 2D model, these two contact points are merged into a single one, and tangential components of the contact forces are present. An important consequence of the absence of tangential forces in the 3D model is that the knit cannot be in equilibrium at zero force.
Indeed, the derivatives of the elastic energy, $\partial \mathcal{E}/\partial \ell_x$ and $\partial \mathcal{E}/\partial \ell_y$, are never simultaneously zero on the energy landscape shown in Fig.~\ref{fig:energy_Landscape}(a).

Finally, the description of elasticity is different. In the 3D case, we considered the elastic rod energy including extension, bending, and torsion (although the latter is not relevant in the absence of friction).
In the 2D model, only bending energy was considered.

\subsection{Energy Landscapes of the 3D and 2D Models.\label{sec:comparison_2D_3D}}

\begin{figure}[htbp]
    \centering
    \includegraphics[width=1.0\linewidth]{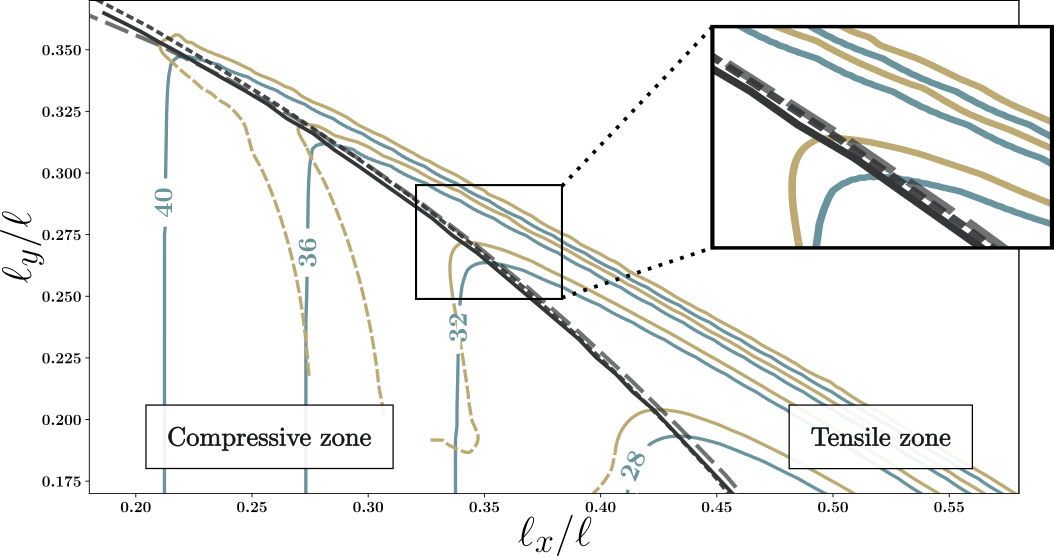}
    \caption{(Colored lines): lines of constant bending energy for the 3D model (blue lines) and for the 2D non-extensible and frictionless model ($\theta_c=0$, yellow lines). 
    Numbers are values of $\mathcal{E}\ell/B$. Dashed parts of the yellow lines correspond to unphysical attractive contact forces $\bff_c \cdot n >0$. 
    (Black lines): line $(f_x=0, \theta_c=0)$ in the 3D model (plain line); line $(f_x=0, \theta_c=0)$ in the 2D model (short dash); line $(f_x=0, f_y=0)$ in the 2D model (long dash). 
    We take $\ell/d=200$.} \label{fig:comparison_2D_3D}
\end{figure}

In the slender limit ($d/\ell \ll 1$), we expect the fabric to be almost planar, and to be accurately described with the 2D model presented in Section~\ref{sec:2D_model}.
To test this assumption, we compare the bending energy of a frictionless ($\mu = 0$) and thin ($d/\ell = 1/200$) fabric computed using the 3D model with that computed using the 2D inextensible model.
Note that, since $\mu = 0$, there is no metastability: for given mesh sizes $(\ell_x, \ell_y)$, the equilibrium solutions of the 2D and 3D models are unique and do not depend on the preparation history. 
This allows for a clean comparison between the outputs of the two models.

For the 3D model, the protocol used to obtain the bending energy is described in Section~\ref{sec:3D-energy}. 
For the 2D model, we solve for regularly spaced values of $\ell_x$ and $\ell_y$ the Type ii) problem (Eq.~\eqref{problem::type2}) with $\{\ell_x,\ell_y\,\theta_c=0\}$ . 
The obtained iso-energy lines are shown in Fig.~\ref{fig:comparison_2D_3D} for both models. The energy contours are in close agreement, supporting the use of the 2D inextensible model for thin-thread fabrics. 
The deviations that appear at small $\ell_x$ and $\ell_y$ arise in regions where the contact forces become attractive in the 2D model ($\mathbf{f}_c \cdot \mathbf{n} > 0$), whereas they remain repulsive in the 3D model. 

\subsection{Energy Valley and Rest-State Line.}

We have just seen that the energy landscapes obtained from the 3D and 2D models for thin frictionless meshes are very similar. These landscapes exhibit an energy valley. The valley can be defined from the energy function $\mathcal{E}(\ell_x,\ell_y)$ as the locus of points satisfying $\nabla \mathcal{E} \cdot \mathbf{e}_{\lambda}=0$~\cite{Eberly1994}, where $\mathbf{e}_{\lambda}$ denotes the eigenvector of the Hessian $(\partial^2 \mathcal{E}/\partial \ell_i \partial \ell_j)$ associated with the direction of positive curvature.
However, this definition does not have an obvious mechanical interpretation. For our purposes, we instead define the valley as the locus of points satisfying $\partial\mathcal{E}/\partial \ell_x=0$. 
This is the line shown in Fig.~\ref{fig:energy_Landscape}(a). 

For the 2D problem, this line may be obtained directly by solving the {\it elastica} problem defined as Type v) (Eq.~\eqref{problem::type5}). 
For different values of $s_c$, we solve:
\begin{equation}
\{f_x=0,\theta_c=0,s_c\} \rightarrow  \{\ell_x,\ell_y,f_y\}.
\end{equation}
The obtained parametric line $\bigl(\ell_x(s_c),\ell_y(s_c)\bigr)$ is shown in Fig.~\ref{fig:comparison_2D_3D}. We observe that it coincides well with the $f_x=0$ line obtained with the 3D model. 
This confirms that modeling the yarns using a planar model provides an excellent approximation of the full 3D model.

We may also compare this line with the line of rest states defined and discussed in Section~\ref{subsec:exp_rest_state}, corresponding to the solution of a Type iii) (Eq.~\eqref{problem::type3}). 
For different values of $s_c$, we solve:
\begin{equation}
\{f_x=0,f_y=0,s_c\} \rightarrow  \{\ell_x,\ell_y,\theta_c\}.
\end{equation}

The obtained line is shown in Fig.~\ref{fig:comparison_2D_3D}. It is almost indistinguishable from the line obtained with $\{f_x=0,\theta_c=0,s_c\}$. 
In other words, when $f_x=0$, imposing $f_y=0$ with $\theta_c\neq0$, or imposing $\theta_c=0$ with $f_y\neq0$, leads to very similar {\it elastica} shapes.
This result is not obvious \emph{a priori}, but may be interpreted as follows. 
When $f_x=0$, the contact force is $\bff_c =f_i \bfe_x + f_y \bfe_y$. 
The moment $M(s)\bfe_z$ of $\bff_c$ at the abscissa $s$ of the centerline is
\begin{equation}
M(s)=f_i\times\left[y(s)-y(s_c)\right]+f_y\times\left[x(s)-x(s_c)\right].
\end{equation}
Since the portion of {\it elastica} between $s_c$ and $\ell/4$ is almost vertical, we have $\vert y(s)-y(s_c)\vert \gg \vert x(s)-x(s_c)\vert$. 
Consequently, forces $f_y$ of the same order of magnitude as $f_i$ can be applied without significantly modifying the bending moment and therefore the curvature of the {\it elastica}.
Applying $f_y$ rotates the contact force (in the sense that it changes the angle $\theta_c$) so that it becomes parallel to the normal vector. 
Consequently, $\theta_c$ can be canceled by applying a force $f_y$ without significantly altering the shape of the {\it elastica}.

\subsection{Knits Made of Thick Threads: $d/\ell \not\simeq0$}

\begin{figure}[htbp]
    \centering
    \includegraphics[width=1.0\linewidth]{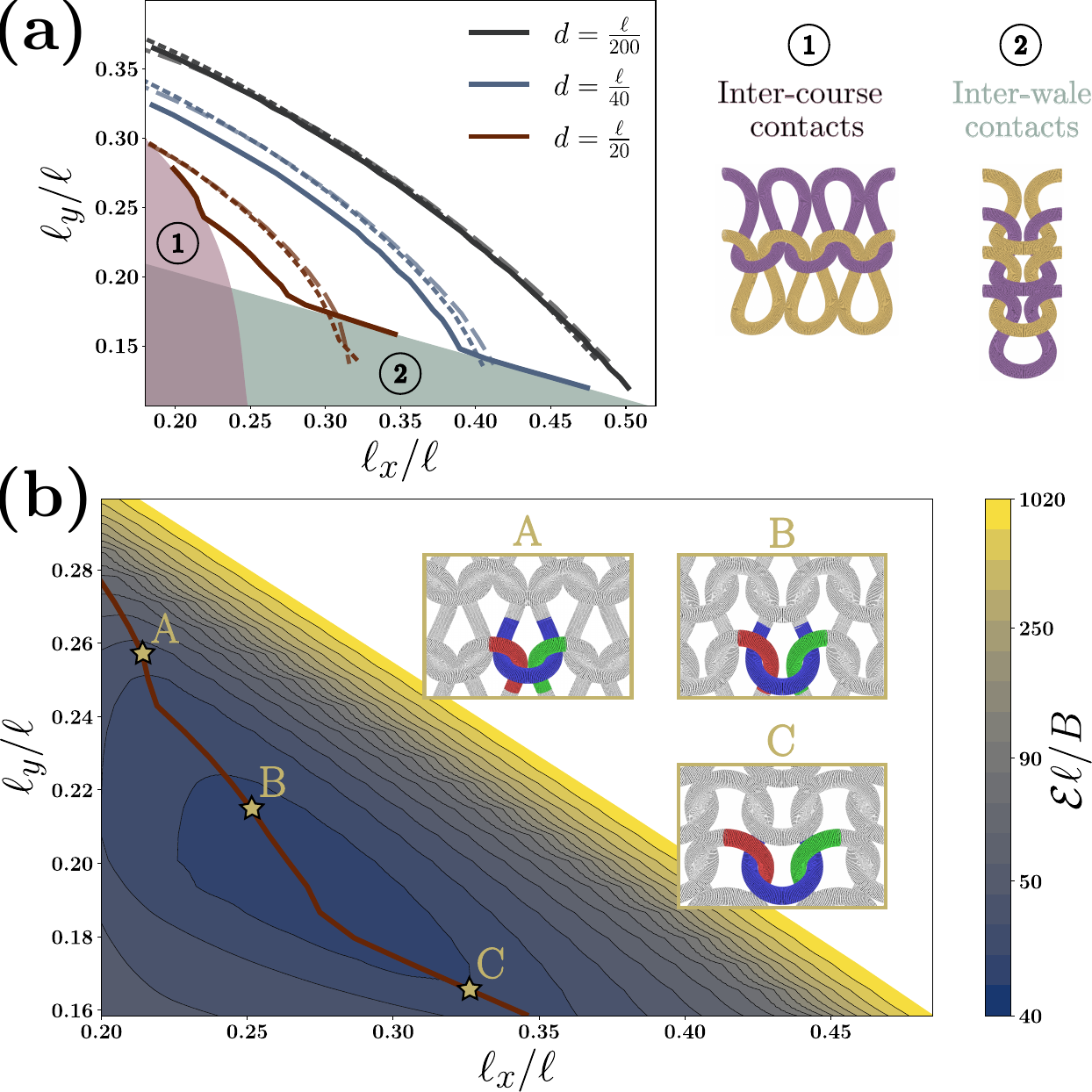}
    \caption{(a) (Plain lines): line $(f_x=0, \theta_c=0)$ of the 3D model; (short dashed line): line $(f_x=0, \theta_c=0)$ of the 2D model; (long dashed line) line \textbf{$(f_x=0, f_y=0)$} of the 2D model. 
    Line colors correspond to different values of $\ell/d$. 
    (Color zones) Zones $1$ and $2$ correspond to regions where inter-course contacts or inter-wales contacts occur. 
    These zones are determined with $\ell/d=20$. 
    (b) Elastic energy $\mathcal{E}$ obtained numerically using the 3D model, with $\ell/d=20$ and $\mu=0$, as a function of the mesh size. (Plain line): line $(f_x=0, \theta_c=0)$. 
    Insets $(A-B-C)$ show the threads in a mesh for different mesh sizes. }
    \label{fig:neutral_lines_vary_d}
\end{figure}

The final point we address is the qualitative and quantitative limits of the two-dimensional model.
To this end, we investigated knitted structures with $\ell/d = 200$, $40$, and $20$. 
For each value, we determined: {\it i}) the $(f_x=0,f_y\ne 0,\mu=0)$ line of the 3D model (the valley of the 3D energy landscape); {\it ii}) the corresponding line of the 2D model (the valley of the 2D energy landscape); and {\it iii}) the $(f_x=0, f_y=0,\mu\neq 0)$ line of the 2D model (the equilibrium line). 
The results obtained with the three models are shown in Fig.~\ref{fig:neutral_lines_vary_d}a.

For $\ell/d=200$, as already discussed in the previous section, these three lines are virtually indistinguishable within the accuracy of their determination. For all values of $\ell/d$ considered (200, 40, and 20), the $(f_x=0, \mu=0)$ and $(f_x=0,f_y=0)$ lines remain very close to one another.
This demonstrates that the approximation introduced in the previous section, consisting of neglecting the influence of $f_y$ on the {\it elastica} shape, remains valid even for $\ell/d=20$. 
By contrast, the energy valleys predicted by the 2D and 3D models progressively diverge as $\ell/d$ decreases. These discrepancies between the 2D and 3D models arise from two distinct reasons.

The first reason is essentially qualitative and does not alter the overall shape of either the $(f_x=0, \mu=0)$ line or the $(f_x=0,f_y=0)$ lines. 
For instance, for $\ell/d=40$, the main difference between the 2D and 3D predictions is a roughly uniform shift of the lines, while the shape of the curves remains unchanged. 
The origin of this shift is straightforward. In the 3D geometry, the out-of-plane displacement of the yarns requires an excess fiber length compared with trajectories that remain strictly planar.
Consequently, for a fixed yarn length $\ell$ per stitch, the projected yarn length in the $z=0$ plane is reduced. As a result, the stitch dimensions that minimize the bending energy must also be smaller. 
This shift is also observed for $\ell/d=20$, where it becomes even more pronounced.

In addition to this global shift, a qualitative change in behavior is observed at small values of $\ell_y$ (for $\ell/d=40$ and $20$) and at small values of $\ell_x$ (for $\ell/d=20$). 
In these regions, the $(f_x=0, \mu=0)$ line predicted by the 3D model departs abruptly from the trend observed at intermediate values of $\ell_x$ and $\ell_y$. 
These deviations originate from the appearance of additional yarn contacts that are absent in a loose knit. 
Indeed, in such a knit, the yarn of row $n$ is in contact only with the yarns of rows $n-1$ and $n+1$. 
As the fabric becomes denser (smaller values of $\ell/d$), additional contacts appear. 
At small values of $\ell_x$, contacts between neighboring loops within the same row $n$ must be considered. 
Such a configuration is illustrated in Fig.~\ref{fig:neutral_lines_vary_d}(a) and corresponds to the contacts occurring in region 1. 
At small values of $\ell_y$, contacts develop between yarns belonging to rows $n$ and $n+2$, corresponding to region 2 in Fig.~\ref{fig:neutral_lines_vary_d}(a).

Since the 2D model only accounts for contacts between successive rows, it cannot reproduce the modifications of the energy landscape induced by these additional steric interactions. 
It is worth noting that for $\ell/d=20$, the region where neither same-row contacts $n$ --- $n$, nor second-neighbor contacts $n$ --- $n+2$, occur becomes narrow. Tight knits where stitches have to contact neighboring ones are called \textit{jammed} and have been identified to also emerge for $\ell/d$ around $16 \sim 20$~\cite{postle1967analysisI,shanahan1973jamming,hepworth1978biaxial}. Finally, the energy landscape (Fig.~\ref{fig:neutral_lines_vary_d}(b)) reveals the emergence of a local minimum in the region without self-contact, associated with the steric interactions between the yarns of rows $n$ and $n+2$.

\section{Discussion and Conclusion.}

This study leverages numerical and analytical models of knitted unit-cells to build a comprehensive mechanical framework based on energy landscapes in mesh-size space. 
The main contributions of the study are the following:
\begin{enumerate}[\label=(\roman*)]
    \item Identification and scaling for the bending- and stretching-dominated responses, with a criterion for the transition.
    \item Expose the role of friction in determining the ensemble of accessible states at rest and under uniaxial external forces, and in altering the response to uniaxial extensions. 
    \item Characterization of the relevant \textit{elastica} model depending on the mechanical questions, and validation against 3D numerical models in the bending regime, and against experiments without any adjustable parameter. 
\end{enumerate}

This study also highlights the crucial role played by the line of rest states in the mesh size space in the mechanical behavior of knitted stitches. 
This line of rest states is derived analytically using a simple double arc model and a bending energy minimization scheme under the geometrical constraints imposed by contact. 
By mapping the bottom of the energy valley, this line controls the energy gradients and hence the resulting mechanics under external forces. 
In the presence of external forces, the stitch shape is only slightly modified from its rest configuration, rotating around what we coined as the neutral point. 
The existence of this point testifies of the subtle coupling between mesh size, thread trajectory, and sliding at the contact point: changes in curvature imposed by external forces or moments can be compensated by internal sliding of the contact to maintain an identical mesh size. 
Furthermore, under forced uniaxial extension of the mesh, the transverse size adapts so as not to drift too far away from the rest line, until jamming of the stitches (or contact with neighboring cells) becomes unavoidable.

Another important point is that the mechanical behavior of loose knits with monofilament elastic threads is captured quantitatively by simple 2D \textit{elastica} models, including in the presence of dry friction. 
Bending energy and geometrical non-linearities are sufficient for the prediction, given that the current state and history of the frictional contact are carefully considered to point to the right mechanical problem. 
As long as we stay in the bending-dominated and slender regimes, 3D numerical simulations are not necessary.
The limits of these regimes have been clearly identified by quantifying the transition to the stretching-dominated regime and how the stitch slenderness ratio $\ell/d$ modifies the energy landscape.

While the present study focuses on the somewhat idealized case of looser monofilament knits, we hope that the framework will serve as a foundation to explore a more diverse variety of fabrics. 
A tighter limit was explored, but still in the purely elastic and incompressible case. 
In practice, knitting plastic monofilament in tight configuration leads to significant plastic deformation of the thread, shifting the minimum energy configuration of the stitch. 
Moreover, one can avoid high plastic deformations in a tight fabric by using multifilament threads; however, these threads have significant cross-section compressibility, adding a new energy scale to the problem~\cite{gonzalez2025pulling}.
In addition, multifilament threads also break the standard relations between the bending, stretching, and twisting moduli, potentially having profound consequences on the energy distribution between the different modes and the resulting energy landscape. 
Quantifying how these added ingredients modify the framework presented in this study is a promising direction to reach quantitative predictions of the mechanics of knitted fabrics from measurable material and frictional properties. 


\textbf{Acknowledgements.} This work has been supported by Agence Nationale de la Recherche Grant ANR-23-CE30-0015. S.P. acknowledges the financial support provided by JSPS KAKENHI Grant-in-Aid for Early-Career Scientists No. 25K17363. J.C. thanks L. Guerin for fruitful discussion on symmetry groups. A.S. and A.F. thank D. Le Tourneau for building the tensile machine, P. Metz and V. Coullon for its control program. 

\section*{Appendix: Circular-arc Model.}

\subsection*{Circular-arc Approximation.}

\begin{figure}[]
    \centering
    \includegraphics[width=0.5\linewidth]{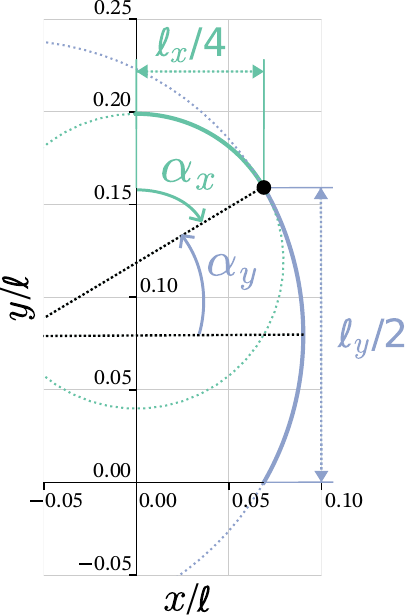}
    \caption{Stitch geometry approximated by two circular arcs with tangent continuity constraint at $s=s_c$ (black dot). Here the stitch is drawn with $(s_c=\ell/12,\ell_x=\sqrt{3}/(2\pi),\ell_y=1/\pi)$.}
    \label{fig_appendix}
\end{figure}

From observation of the shape of the stitches given by the 2D \textit{elastica} model (see Fig.~\ref{fig:lxly_full}(b) and Fig.~\ref{fig:lxly_fixed_forces}(b), for instance), a reasonable approximation is to divide the stitch at the contact point into two circular arcs of constant curvature.
The first arc runs between $s\in[0,s_c]$ and has a chord of length $\ell_x/4$, while the second runs between $s\in[s_c,\ell/4]$ and has a chord of length $\ell_y/2$. 
We note $\kappa_x$ and $\kappa_y$ the curvatures of the arcs with chords of lengths $\ell_x/4$ and $\ell_y/2$, respectively. 
We define $\alpha_x$ and $\alpha_y$ as the opening angles of the arcs, such that the first arc runs between $\alpha\in[0,\alpha_x]$, and the second one runs between $\alpha\in[-\alpha_y,\alpha_y]$.
The curvatures, opening angles, and arc lengths are directly related by:
\begin{subequations}
\begin{align}
    \alpha_x &=\kappa_x s_c,
    \label{eq:alphax} \\
    \alpha_y &= \kappa_y(\ell/4-s_c)/2. \label{eq:alphay}
\end{align}
\end{subequations}
The chords of the arcs are related to these quantities by the following relations:
\begin{subequations}
\begin{align}
    \sin\alpha_x&=\sin(\kappa_x s_c)=\frac{\ell_x\kappa_x}{4}, 
    \label{eq:cordx} \\
    \sin\alpha_y&=\sin(\kappa_y(\ell/4-s_c)/2)=\frac{\ell_y\kappa_y}{4}. \label{eq:cordy}
\end{align}
\end{subequations}
From these curvatures, the bending energy of the stitch directly writes: 
\begin{equation}
    \mathcal{E}_{\mathrm{bend}}^{arc} = \frac{B}{2}\left({ s_c \kappa_x^2+(\ell/4-s_c)\kappa_y^2}\right).
\end{equation}
The tangent continuity condition at the contact writes $\alpha_x+\alpha_y=\pi/2$ and may be expressed as a constrain $\mathcal C=0$ on curvatures:
\begin{equation}
    \mathcal C(\kappa_x,\kappa_y,s_c)=s_c \kappa_x + (\ell/4 - s_c) \kappa_y /2 -\pi/2 =0.
    ~\label{eq:constrain:c}
\end{equation}

\subsection*{Analytical Derivation of the Rest-State Line.}

From these expressions, we can derive the parametric equation of the line $(f_x=0, f_y=0)$. The total potential energy to be minimized is:
\begin{equation}
    \begin{split}
\mathcal{L}(\kappa_x,\kappa_y,s_c)&=\mathcal{E}^{arc}_{bend}(\kappa_x,\kappa_y,sc)\\
&- M(0)\Theta(0) - M(s_c)\Theta(s_c) - M(\ell/4)\Theta(\ell/4)\\
&+\lambda \mathcal C(\kappa_x,\kappa_y,s_c).
    \end{split}
\end{equation}
The first term is the bending energy. 
The three terms $-M(s)\Theta(s)$ are the potential energy due to the moments applied at $s=0$, $s=s_c$, and $s=\ell_4$, with $\Theta(s)$ the local slope with respect to $\bfe_x$. 
The last term is the constraint $\mathcal C$, with $\lambda$ the associated Lagrange multiplier. 
The moments at the ends satisfy $M(0)=B\kappa_x$, $M(\ell/4)=-B\kappa_y$, and the equilibrium condition writes $M(0)+M(s_c)+M(\ell/4)=0$. 
The angles are $\Theta(0)=0$, $\Theta(s_c)=-\alpha_x$, and $\Theta(\ell/4)=-\pi+\alpha_x$. 
The total potential energy is then:
\begin{equation}
\mathcal{L}(\kappa_x,\kappa_y,s_c) =-\frac{B}{2}\left({ s_c \kappa_x^2+(\ell/4-s_c)\kappa_y^2}\right)
+\lambda \mathcal C(\kappa_x,\kappa_y,s_c).
\end{equation}
For any fixed value of $s_c$, we seek the values of the curvatures that minimize $\mathcal{L}$:

\begin{subequations}
\begin{align}
\frac{\partial \mathcal{L}}{\partial \kappa_x}|_{\kappa_y,s_c}&=0,\\
\frac{\partial \mathcal{L}}{\partial \kappa_y}|_{\kappa_x,s_c}&=0.
\end{align}
\end{subequations}

Eliminating $\lambda$ from these equations gives:
\begin{equation}
    \kappa_x= 2\kappa_y. \label{eq:app:kappaxy}\\
\end{equation}

With this circular-arc approximation, the configurations that minimize the bending energy for a given contact position $s_c$ are simply those for which the curvature in the $x$-direction is twice that in the $y$-direction.
Substituting this relation in the $\mathcal {C} = 0$ condition gives the explicit expression for the curvatures as a function of $s_c$, 
\begin{subequations}
    \begin{align}
\kappa_x&=2 \pi/(3s_c/2 + \ell/8), \label{eq:app:kappax}\\
\kappa_y&=\pi/(3s_c/2 + \ell/8). \label{eq:app:kappay}
\end{align}
\end{subequations}
From Eqs.~\eqref{eq:cordx} and \eqref{eq:cordy}, the explicit expression of the minimal bending energy line is: 
\begin{subequations}
\begin{align}
    \ell_x^{arc}&=\frac{(6 s_c +\ell/2)}{\pi}\sin\Bigl(\frac{\pi s_c}{3s_c/2+\ell/8}\Bigr), \label{eq:app:ellx}\\
    \ell_y^{arc}&=\frac{2(6 s_c +\ell/2)}{\pi}\cos\Bigl(\frac{\pi s_c}{3s_c/2+\ell/8}\Bigr). \label{eq:app:elly}
\end{align}
This line is drawn in Figure~\ref{fig:lxly_full}(a) alongside the prediction of the 2D \textit{elastica} model.
\end{subequations}

\subsection*{Analytical Derivation of the Neutral Point $N$.\label{subsec:app:arc:neutral}}

If extra forces $f_x$ or $f_y$ are applied, the potential energy due to the external forces must be added, and the total potential energy to be minimized becomes:

\begin{equation}
\begin{split}
\mathcal{L}(\kappa_x,\kappa_y,s_c)&=\mathcal{E}^{arc}_{bend}(\kappa_x,\kappa_y,s_c)\\
&- M(0)\Theta(0) - M(s_c)\Theta(s_c) - M(\ell/4)\Theta(\ell/4)\\
&- f_x \bfr(0) - f_c \bfr(s_c) -f_y \bfr(\ell/4)\\
&+\lambda \mathcal C(\kappa_x,\kappa_y,s_c).
\end{split}
\end{equation}

The minimization of $\mathcal{L}$ with respect to $\kappa_x$ and $\kappa_y$ at fixed $s_c$ shows that the condition $\kappa_x = 2\kappa_y$ (Eq.~~\eqref{eq:app:kappaxy}) no longer holds. 
However, the continuity constraint (Eq.~~\eqref{eq:constrain:c}) and the geometry constraints (Eqs.~\eqref{eq:cordx} and \eqref{eq:cordy}) remain valid.

In the following, we propose an analytical derivation of the coordinates of the neutral point $N$. The neutral point lies on the line of rest states without external forces. To determine its coordinates, we search along the rest-state line for a configuration that satisfies the defining property of the neutral point. Let $\ell_x(s_c,\kappa_x,\kappa_y)$ and $\ell_y(s_c,\kappa_x,\kappa_y)$ be the mesh sizes of a rest state with a contact point at abscissa $s_c$. By considering small variations of $s_c$, $\kappa_x$, and $\kappa_y$ around this state, the cell sizes become:

\begin{subequations}
    \begin{align}
    &\ell_x(s_c+\Delta s_c,\kappa_x + \Delta\kappa_x,\kappa_y+\Delta\kappa_y),
    \label{eq:app:deltalx} \\
    &\ell_y(s_c+\Delta s_c,\kappa_x + \Delta\kappa_x,\kappa_y+\Delta\kappa_y).
    \label{eq:app:deltaly}
\end{align}
\end{subequations}
For small variations, we can further write:
\begin{subequations}
    \begin{align}
    \Delta \ell_x =& \frac{\partial \ell_x}{\partial s_c}|_{\kappa_x,\kappa_y}\Delta s_c
        + \frac{\partial \ell_x}{\partial \kappa_x}|_{s_c,\kappa_y}\Delta \kappa_x
        + \frac{\partial \ell_x}{\partial \kappa_y}|_{s_c,\kappa_x}\Delta \kappa_y,\\
     \Delta \ell_y =& \frac{\partial \ell_x}{\partial s_c}|_{\kappa_x,\kappa_y}\Delta s_c
        + \frac{\partial \ell_x}{\partial \kappa_x}|_{s_c,\kappa_y}\Delta \kappa_x
        + \frac{\partial \ell_x}{\partial \kappa_y}|_{s_c,\kappa_x}\Delta \kappa_y.
\end{align}
\end{subequations}
Taking into account that $\frac{\partial \ell_x}{\partial \kappa_y}|_{s_c,\kappa_x}=\frac{\partial \ell_y}{\partial \kappa_x}|_{s_c,\kappa_y}=0$ (see Eqs.~\eqref{eq:cordx} and \eqref{eq:cordy}), and that $s_c \Delta \kappa_x + (\ell/4 - s_c) \Delta \kappa_y=0$ (constraint $\mathcal{C}=0$, Eq.~\eqref{eq:constrain:c}), the equations for $\Delta \ell_x$ and $\Delta \ell_y$ become:
\begin{subequations}
\begin{align}
    &\Delta \ell_x = &\frac{\partial \ell_x}{\partial s_c}|_{\kappa_x,\kappa_y}\Delta s_c 
        &+ &\frac{\partial \ell_x}{\partial \kappa_x}|_{s_c,\kappa_y}\Delta \kappa_x,&\\
    &\Delta \ell_y = &\frac{\partial \ell_y}{\partial s_c}|_{\kappa_x,\kappa_y}\Delta s_c 
        &+ &\frac{\partial \ell_y}{\partial \kappa_y}|_{s_c,\kappa_x}\Delta \kappa_x &\times \Bigl(\frac{-s_c}{(\ell/4 - s_c)/2}\Bigr).\qquad
\end{align}
\end{subequations}
This system may be written in matrix form:
\begin{equation}
\begin{pmatrix}
\Delta \ell_x\\
\Delta \ell_y
\end{pmatrix}=	M \begin{pmatrix}
\Delta s_c\\
\Delta \kappa_x
\end{pmatrix},
\end{equation}
with $M$ given by the preceding equations. 
This equation relates the variations in cell sizes to the variations in $\kappa_x$ and $s_c$. 
The cell sizes may remain constant ($\Delta \ell_x = \Delta \ell_y = 0$) for finite variations of $\Delta \kappa_x$ and $\Delta s_c$ if: 
\begin{equation}
    \det(M)=0.
\end{equation}
The different terms of $M$ can be expressed as a function of $\kappa_x$, $\kappa_y$, $\alpha_x$ and  $\alpha_y$. The condition $\det(M)=0$ becomes:
\begin{equation}
    F(\alpha_x) /2 \kappa_x = F(\alpha_y) / \kappa_y,
\end{equation}
with $F(\alpha)=\tan(\alpha)/\alpha - 1$. 
Using the fact that, on rest-state line, $\kappa_x=2 \kappa_y$, and that $\alpha_y=\pi/2-\alpha_x$, we obtain: 
\begin{equation}
    F(\alpha_x) = 4 F(\pi/2 - \alpha_x)
\end{equation}
Solving this equation numerically gives $\alpha_x= 0.9851..$. 
From the value of $\alpha_x$ at the neutral point, together with the fact that this point lies on the rest-state line and must satisfy Eqs.~\eqref{eq:app:kappaxy}, \eqref{eq:app:kappax}, \eqref{eq:app:kappay}, \eqref{eq:app:ellx} and \eqref{eq:app:elly}, it follows that:
\begin{subequations}
    \begin{align}
s_c^N/\ell &= 0.0740..\\
\ell_x^N/\ell &= 0.2504.. \\
\ell_y^N/\ell &= 0.3322..
    \end{align}
\end{subequations}
These values define the neutral point $N^\mathrm{arc}$, whose position is shown in Fig.~\ref{fig:lxly_fixed_forces}(a). It lies very close to the neutral point estimated from the full 2D \textit{elastica} model.

\bibliographystyle{unsrt}
\bibliography{reference.bib}

\end{document}